\documentclass[12pt]{article} 
\usepackage{graphicx}
\usepackage{multicol}
\usepackage{color}
\definecolor{verdon}{cmyk}{1,0.5,1,0}
\definecolor{blue}{cmyk}{1,1,0,0}
\definecolor{red}{cmyk}{0.2,1,1,0.0}
\newcommand{\cita}[1]{{\color{blue} \cite{#1}}}
\begin{document}
\small

\title{\color{verdon} Is there a problem with low energy SN1987A neutrinos?}

\author{
M L Costantini$^{1,2}$, A Ianni$^1$, G Pagliaroli$^{1,2}$, F Vissani$^{1}$\\
$^1${\small\em INFN, Laboratori Nazionali del Gran Sasso, Assergi (AQ), Italy}\\
$^2${\small\em University of L'Aquila, Coppito (AQ), Italy}}

\date{}

%\ead{marialaura.costantini@aquila.infn.it}

%\author{
%\address{INFN, Laboratori Nazionali 
%del Gran Sasso, Assergi (AQ), Italy}
%%\ead{aldo.ianni@lngs.infn.it}

%\author{G Pagliaroli}
%\address{INFN, Laboratori Nazionali 
%del Gran Sasso, Assergi (AQ) and University of L'Aquila,
%Coppito (AQ), Italy}
%\ead{giulia.pagliaroli@lngs.infn.it}

%\author{F Vissani}

%\address{INFN, Laboratori Nazionali 
%del Gran Sasso, Assergi (AQ), Italy}
%\ead{francesco.vissani@lngs.infn.it}

\maketitle

\def\abstractname{\color{red}\bf Abstract}
\begin{abstract}
{\footnotesize
The observation of several low energy events during SN1987A burst 
has an important weight in the attempts to learn about the properties of 
the supernova neutrinos, but is somewhat puzzling if compared with  
the expectations. 
In this work, we study the low energy events observed by Kamiokande-II,
and consider in particular 
the possibility that a few of them are due to background.
We focus our attention not only 
on the event 6, that falls below the energy threshold of 7.5 MeV, 
but also on the other four events  that at 1 $\sigma$ fall 
below the energy threshold, namely the events 3,4,10,12.
The {\em a priori} expectations on the number of background events 
above threshold would suggest 1 or 2 background events only. 
The volume distribution of the Kamiokande-II events is not a uniform 
distribution at 3 $\sigma$, that suggests the presence of background events 
close or at the border of the volume used for the analysis, 
including the events 3,4,10. 
Next we checked the expected energy distribution assuming that the signal is 
due to $\bar{\nu}_e p\to e^+ n$ and that the average antineutrino energy is 
14 MeV. The agreement with the observations is not perfect,
but it is acceptable at the 11 \% confidence level 
if we include the peak of low energy background events; 
otherwise, we face a 2.9~$\sigma$ problem.
The expected energy distribution 
implies that the evidence for supernova neutrinos is at 10 $\sigma$
and that 1-3 background events are plausible. 
This conclusion does not change strongly when we model 
the time distribution of the signal, taking into account
the presence of an initial luminous phase of neutrino emission.
This suggests however that 
some of the early events (in particular, event 4) are 
due to supernova neutrinos and not to background.
In summary, our comparison between the expectations and the data 
lead us to formulate the hypothesis that some of the observed 
low energy events are due to background and that some among 
them belong to a peculiar phase of emission, 
that could be further characterized by low energy neutrinos.
Such an interpretation 
is mostly attractive, since it diminishes to a minimum the postulated 
number of background events and thus improves the agreement between 
the {\em a priori} (model independent) and the {\em a posteriori} 
(model dependent) expectations on the number of background events. 
We argue on these grounds that there is no 
significant disagreement between the average 
energy of the supernova neutrinos 
seen in Kamiokande-II and the conventional expectations. 
Alternative possibilities of 
interpretations are mentioned and briefly discussed. }
\end{abstract}

{\footnotesize 
\def\contentsname{\centerline{{\small\bf\color{red}  Contents}}}
\sf \tableofcontents}

%\pacs{97.60.Bw, 95.85.Ry, 95.55.Vj}

\section{\sf\color{verdon}  Motivation and context \label{abh}}
On February 23, 1987, several experiments~\cita{hirataPRL,imb,baksan,mb}
contributed to 
begin the era of extragalactic neutrino astronomy. 
These observations had an enormous impact on astrophysics and
on particle physics.\footnote{In order 
to have an idea of how large was the impact
it is sufficient to consider that, on March 2007, 
there are 172,000 (resp., 3,050) entries on `SN1987A' 
in {\sc Google} (resp., in {\sc Google scholar}). 
The {\sc SPIRES} database lists 578 papers
typing the same keyword;
our first two references are those with
more than 500 citations, 
whereas those that rank more than 100 citations 
include theoretical works on extra dimensions, 
CPT violation, axions, magnetic moments and exotic particles; several 
studies of the role of neutrino oscillations follow.
Finally, the 20th anniversary of this observation was 
celebrated with conferences held in Moscow (Feb.~20-22, 2007), 
in Hawaii (Feb.~23-25, 2007) and Venice (Mar.~6-9, 2007)~\cita{webs}.} 
A straightforward interpretation 
of these data is yet {\em difficult}. 
This is due not only to the fact that 
we lack a firmly established theory of supernova explosion, 
but also to certain anomalous features of the data that have been 
understood, emphasized and analyzed 
in the course of the time. 
In particular, this is the case 
of the average energy deduced by the 12
events observed in Kamiokande-II (KII), 
%, see also 
%references  \cita{hirata,hirataPHD,moriond}), 
which is half of
that observed by IMB, and also lower then what 
expected in theoretical models for supernova 
neutrino emission, recently reviewed in~\cita{keil}.
%(the theoretical works we refer in this paper are 
%\cita{nad,del1,del2,latti,bahcall,keil,ale,del3}).
The main goal of the present paper is to propose 
an interpretation of this apparent discrepancy.

The low energy feature of KII dataset 
is clearly reflected in the outcome of two recent
analyses of these data, that explore certain
possibilities that deviate strongly from 
theoretical expectations.
The first one is by Mirizzi and Raffelt~\cita{mira}, who 
describe the distribution in $\bar{\nu}_e$ energies as
$E_{\bar{\nu}_e}^\alpha \exp[-(\alpha+1) E_{\bar{\nu}_e}/\langle E \rangle]$
and find that the best fit of KII data is provided by $\alpha\sim 0$
({\em i.e.}, a monotonically decreasing distribution).
The second one is the analysis of Lunardini~\cita{luna}, who 
adopts a two-component distribution as suggested
by three flavor oscillation scenarios and finds that a
component with $\langle E_{\bar{\nu}_e} \rangle\sim 5$~MeV
permits to fit KII data better.\footnote{In fact, the 
question of which is the energy distribution of an `average' 
core collapse supernova 
has important implications for future experiments: {\em e.g.}, 
the analysis of SN1987A neutrinos of \cita{luna} was used 
to argue that future search of relic supernova neutrinos 
could fail because the emitted neutrinos have 
much lower energy than expected.}
This type of approaches is useful to emphasize 
features of the data like the  
excess of low energy events in KII dataset
but one should recall that neither 
$\langle E_{\bar{\nu}_e} \rangle\sim 5$ MeV nor 
$\alpha\sim 0$ are compatible with the current 
expectations for supernova neutrino emission,
so that in a conservative analysis, 
these values should not be allowed.

Stated otherwise, there is a trend 
in modern studies of SN1987A to accept the opinion 
that KII data are incompatible with the 
present theoretical expectations and, consequently, 
to start from this position to investigate more or less 
radical departures from the conventional 
paradigm for neutrino emission. One could even 
think that the failures to obtain a theory of supernova 
explosion could motivate such an attitude. 
But the fact that we do not have yet a definitive theory
of the explosion does not mean that all theoretical possibilities
are {\em a priori} equivalent. Furthermore, 
one should be careful in distinguishing between 
the problems of getting a theory of supernova explosion 
and the problem of knowing the distribution of 
the emitted neutrinos. In the only scenario 
that has been explored in some details till now, the 
so called ``delayed scenario'' \cita{del1,del2}, a large amount 
of the neutrinos--up to 90~\%--is emitted in a phase that 
{\em follows} the explosion: thus, the two problems are 
to a certain extent independent (we will better 
analyze the connection in the following). 
Comparing~\cita{keil} and~\cita{bahcall} we note that 
the expected range of one crucial parameter, the average $\bar{\nu}_e$ 
energies or temperatures,  did not change much since 1989.
What changed is the expectation on the  
temperature of the other antineutrinos 
and thus the impact of oscillations, that 
after~\cita{keil} can be argued  to modify only slightly 
the observable $\bar{\nu}_e$ signal. To summarize, the main reasons 
why we are not convinced that we should abandon the conventional paradigm
for neutrino emission are that: 
1)~the expectations for neutrino emission seem to be stable;
2)~conversely, there is no convincing theoretical argument till now  
supporting an interpretation of the excess of low energy events 
in terms of low energy supernova neutrinos.

Also, on general statistical ground one should be aware of the 
risks of using the data--12 events in the case of KII--to 
infer the characteristics of the model, rather than asking 
the significance level at which the null-hypothesis (=the theoretical
expectation) is ruled out. 
In principle it is always possible to obtain a perfect 
fit ($\chi^2=0$) to the data declaring that the observed 
distribution coincides with the expected one, but only
if one is ready to renounce to the previous knowledge.
In short, we believe that a \underline{conservative} 
discussion should address other questions: 
How severe is the deviation from the conventional theoretical expectations? 
How reliable is the indication of a large amount of low 
energy supernova neutrinos, inferred from KII observations? 
Is it possible to conceive other interpretations 
(more standard than the one proposed in \cita{luna} and \cita{mira}) 
of the observed low energy events?

This is why we would like to explore the
possibility that some of the low energy events in
KII are not due to $\bar{\nu}_e p\to e^+ n$ interactions
of supernova neutrinos as usually assumed. 
The possibility that a few events are due to
elastic scattering has been recently reconsidered~\cita{aldo},
finding that, although this cannot be excluded, it
helps only marginally to explain the excess of low
energy events in KII data set.
This forced the authors of~\cita{aldo} to admit
that a few of these events are of
a different origin, and possibly are 
due to background. This is a conservative 
position, since it is evident that the KII observations 
could be polluted by some background events: 
see figures~4, 5, 10, 11, 12 of the second 
publication of KII collaboration on SN1987A, {\em i.e.}~\cita{hirata}.
Furthermore, in \v{C}erenkov type detector it is not possible to know 
whether a single event is due to signal or to background.
%for instance, even in the first KII publication,  \cita{hirataPRL}, 
%the analysis is introduced
%by the  sentence 
%\begin{quote}
%{\em ``assuming that nine of the twelve events are due to
%$\bar{\nu}_e p^+\to e^+ n$}''. 
%\end{quote}
In our understanding, the 
main point of KII collaboration was just that it is not possible 
that all 12 events are due to background, which means that 
an observation of supernova neutrinos has been made
(as we show in Sect.~\ref{ed}, the {\em a priori} 
knowledge of the signal reinforces significantly this evidence).
It should be noted that the KII collaboration did not quantify 
the probability that the 12 events are due to background 
on individual basis; the only step in this direction 
was to exclude occasionally
one specific event, occurred 
below the threshold for solar neutrinos events 
(more discussion on this issue in Sect.~\ref{et}).
In this note, we analyze quantitatively the possibility 
that some background events occurred in the KII dataset,
extending and complementing the analysis of~\cita{lambLoredo}.
In particular, we focus our attention
on the spatial and the energy distribution of the 12 candidate events.

We are aware of the problems of small numbers statistics, that 
there are many more papers on SN1987A neutrinos than events, and that 
there is the risk of running into `forensics' (quoting a witticism 
of John Beacom). 
Nevertheless, we should also recall that supernovae are 
rare events on human timescale, and that all we have at 
the moment is one observation of supernova neutrinos and 
some theoretical ideas to compare with. In other words, we 
feel that we have the duty to analyze all possible hints for 
anomalies and to extract as much information as we can 
from SN1987A neutrinos.

The plan of this paper is the following: 
in Sect.~\ref{sep} we discuss the tools we have 
to separate signal from background events; 
in Sect.~\ref{discu} we formulate definite hypotheses on
the background and on the signal events, and show that the
average energy of SN1987A neutrinos 
agrees much better with the expectations when we
account for the presence of a few 
background events; in Sect.~\ref{ddu} we comment 
on the approach we used and the results we obtained, and 
finally draw our conclusions.

\section{\sf\color{verdon}  Tools to separate 
signal from background in Kamiokande-II dataset\label{sep}}
We begin recalling what are the known and the unknown aspects 
for the analysis of the events observed in KII. 
Several (though not all) characteristics of the background are 
known: the events are Poisson distributed with 
a given rate, they have relatively low energy, and 
they are more frequent in the border 
of the volume used in the analysis~\cita{hirata},
that touches the planes of the photo-tubes.
The characteristics of the signal are, instead, 
known theoretically: the events are distributed 
uniformly in the detector, they have relatively 
high energy, and their rate is not precisely known.
Going into details, the characteristics 
of the background and of the signal 
that we will use are:

{\em 1)} 
The volume of the KII detector used in the analysis of SN1987A neutrinos
has 
\begin{equation}
\mbox{radius of }R=7.2\mbox{ m, total height of }H=13.1\mbox{ m,} 
\end{equation}
for a volume of 2140 tons.
This volume is much larger than the 680 tons 
fiducial volume defined in~\cita{hirataPHD},
where background events are rare.
We know that this wide volume is not free from background~\cita{hirata}.
The energy distribution of the background in our figure~\ref{fig2} 
is obtained by multiplying the background per unit time 
given in figure~2a of Lamb and Loredo~\cita{lambloredoPRIV}
by the time elapsed from the first to the last events, 
\begin{equation}
T=12.439\mbox{ sec}
\label{times}
\end{equation}
that makes 2.3 background events on average and 0.272 
of them above the threshold of 7.5 MeV. 
The assumption on the background has been validated 
by checking that the distributions 
of $N_{hit}$ obtained from figure~4 of~\cita{hirata} 
and the average number of background events shown in figure~10 
of the same KII paper are in good agreement with what we obtain with 
our distribution. More discussion of the assumed background is in 
Sect.~\ref{et} and  App.~\ref{app:aa}.

{\em 2)} 
The supernova neutrino signal is supposed to be due to
$\bar\nu_e p\to n e^+$ (`inverse beta decay') reaction.
We use the cross section in~\cita{ale} to perform 
the integration on the neutrino energy using 
the full matrix element and kinematical range.
The antineutrino flux that we assume is 
the one described in Sects.~I and~II of~\cita{aldo},
with the parameters of equation~8 there
($\bar{\nu}_\mu$ and $\bar{\nu}_\tau$ average energy 10~\% higher
than $\bar{\nu}_e$ as in~\cita{keil}, energy equipartition and
$\alpha=3$), including oscillations. 
The two crucial input parameters are:
\begin{equation}
\begin{array}{rcl}
\mbox{energy radiated in }\overline{\nu}_e&=& 
4\times 10^{52} \mbox{ erg}\\[.5ex]
\mbox{average energy of }\overline{\nu}_e&=& 14 \mbox{ MeV}
\end{array}
\label{parama}
\end{equation}
The first value is just in the middle of theoretical 
expectations. Instead, the value of the average  
energy of $\overline{\nu}_e$ is on the low side,
but compatible with  
the present theoretical expectations \cita{keil}. 
Furthermore, it fits well the IMB 
observations \cita{imb},
that consisting of a sample of 8 high energy events
are in practice background free. 
Two simple tests permit us to check that 
these values are reasonable:
a)~When the emitted antineutrino energy is 
compared with the (fermionic) Stefan-Boltzmann law 
$L_{\bar\nu_e}=7\pi^3/240\ R^2\ T^4$, it 
corresponds to an emission from a spherical surface of 
radius $R=12$ km that lasts 12.6 seconds.
b)~The energy radiated 
compares well with Lattimer-Yahil formulae \cita{latti,nadd}.
In fact, assuming the total emitted energy is 6 times larger than the one 
emitted in antineutrinos (equipartition), 
the expected value $E_b=1.5\times 10^{53} (M_g/M_\odot)^2$ erg 
gives $M_g=1.3 M_\odot$; 
then the initial mass of the stellar core 
is $M_c = M_g (1 + 0.084 M_g/M_\odot)=1.4 M_\odot$.
See~\cita{aldo} for further discussion of the selected model.
With this model, we expect 11.8 signal events in Kamiokande-II.

With more detailed experimental information,
more precise statements could be possible;  
it would be useful to know the background rate for any single 
event (taking into account the specific position, number of hit
phototubes and photoelectrons, possible correlations, {\em etc});   
the error on the position of any event;   
the energy distribution of the {\em two} components 
of the background, namely the one which is located in the border 
(of higher energy) and the one that is distributed in the volume
(of lower energy).
With reference to the last point,
we quote the Kamiokande-II 
collaboration~\cita{hirata}: the events
with 
$N_{hit}\ge 23$ ``are consistent with higher-energy products
of radioactivity at or outside the tank wall'' whereas
``the events with $N_{hit}\le 20$ are largely
due to ${}^{214}$Bi decay'' (from the $^{222}$Rn chain)
and thus, presumably, more uniformly
distributed.
In the present paper we resort only to the published 
information on the background \cita{hirata,moriond,hirataPHD,lambLoredo}.

In the following of this section, we will obtain some new 
hints on possible background events 
without the need of adopting a specific model 
for neutrino emission. However, when we will pass to the comparison of the 
expectations and the observed energy distribution, we will 
adopt the well-specified model for the signal of eq.~\ref{parama}
obtaining in this way other hints on possible background events.
At this point, one could think that the 
conclusions that we reach on background are not reliable, since they 
depend at least partly on the assumed model for neutrino emission.  
Thus, in order to avoid confusion, we prefer to repeat 
that the first purpose of our analysis is to check whether there 
is a problem with the conventional expectations for the 
supernova neutrinos, and the second is to ask whether 
this problem can be solved by making reasonable 
assumptions on the background. So, the use of a theoretically 
motivated model to compare with is not only justified, but is in 
fact a necessary first step. We will stick as much as possible to
the very simple model defined above,
but will also consider reasonable (theoretically acceptable)
modifications from the assumed model, showing that 
most of the results and the indications we obtain are stable.
Of course, if 
a new paradigm for neutrino emission will eventually emerge, it will 
be important to update these considerations and repeat 
the calculations described below.

The tools that we describe in this section  
are the spatial  (Sect.~\ref{sd}),  
the energy distribution of the events (Sect.~\ref{ed})
and the energy threshold used by Kamiokande-II
collaboration (Sect.~\ref{et}). The compatibility of the indications
will be discussed later (Sect.~\ref{discu}).

\begin{table}[t]
\begin{center}
{\footnotesize
\begin{tabular}{|c||ccc|ccc|ccc|ccc|}
\hline
 & 1 & 2 & 3 & 4 & 5 & 6 & 7 & 8 & 9 & 10 & 11 & 12 \\ \hline
$N_{hit}$ & 58 & 36 & 25 & 26 & 39 & 16 & 83 & 54 & 51 & 21 & 37 & 24 \\
$E $ [MeV] & 20.0 & 13.5 & 7.5 & 9.2 & 12.8 & 6.3 & 35.4 & 21.0 & 19.8 & 8.6 & 13.0 & 8.9 \\ 
$\delta E$ [MeV] & 2.9 & 3.2 & 2.0 & 2.7 & 2.9 & 1.7 & 8.0 & 4.2 & 3.2 & 2.7 & 2.6 & 1.9 \\
\hline
$\theta_\perp$ [deg] & 14 & 95 & 40 & 36 & 66 & 137 & 56 & 50 & 39 & 51 & 70 & 106 \\
$D_{out}$ [m] & 13 & 0.7 & 16 & 14 & 13 & 2.7 & 13 & 13 & 10 & 0.2 & 10 & 9 \\
$D_{in}$ [cm] 
& 200 & 240 & 13 & 2 & 150 & 1200 & 14 & 5 & 640 & 3 & 450 & 530 \\
$d_{min}$ [cm] & 200 & 9 & 10 & 1 & 70 & 200 & 13 & 3 & 500 & 2 & 300 & 300 \\
$f$ [\%] & 37 & 96 & 96 & 99 & 72 & 35 & 94 & 99 & 3 & 99 & 20 & 21 \\
\hline
\end{tabular}}
\caption{\em Value of certain observables described in the text.
First line, progressive event number; first column, the selected observables.
\label{tab1}}
\end{center}
\end{table}

\subsection{\sf\color{verdon}  Spatial distribution \label{sd}}

A lot of useful information about each of the 12 events
is given in tables~I and II of~\cita{hirata}:
the number of hit photo-tubes $N_{hit}$,
the reconstructed energy of the events (using additional information) and
its uncertainty, the Cartesian coordinates $x_i,y_i,z_i$
(but not the error on the position)
and finally the reconstructed direction of the momentum
$\cos\alpha_i,\cos\beta_i,\cos\gamma_i$.
In this way
we can deduce several interesting quantities,~{\em e.g.},
\begin{enumerate}
\item[$\theta_\perp$] the angle between the direction of the momentum
and the normal to the closest surface of the volume (events
with $\theta_\perp\sim 0$ are presumably seen better);
\item[$D_{out}$] the distance between the coordinate of the event
and a formal `exit point' from the volume,
following the direction of the momentum ($D_{out}$ should be large in
comparison to the distance between photo-tubes, about 1 meter, for
an event to be reliably reconstructed);
\item[$D_{in}$] the distance between the coordinate of the event
and a formal `entry point' in the volume, following again 
the momentum but in the opposite direction;
\item[$d_{min}$] the minimal distance from the border of the volume
used in the analysis (that can warn us against possible background events);
\item[$f$] the volumetric coordinate, namely
the fraction of the volume contained in a concentric
sub-detector of the same shape and with the
same geometrical center of KII, that is 
formally defined 
as follows:
\begin{equation}
f_i=\mbox{max}\left[ \frac{\sqrt{x_i^2+y_i^2}}{R}
\ ,\
\frac{\mbox{abs}(h_i)}{H/2}
\right]^3
\end{equation}
where we set $h_i=z_i-60$ cm, taking into account 
the offset of the origin of the coordinates.
\end{enumerate}
The last quantity is of particular interest, since 
when we increase the volume of the sub-detector, the number of 
supernova neutrino events should increase by the same factor
($i.e.$, $f$ can be considered as the cumulative distribution
coming from a uniform probability distribution function).

In App.~\ref{app:bb} we show that, assuming uniformity
(namely ignoring systematics effects due to the detector
response) we are {\em underestimating} the impact of 
a possible contamination from 
relatively high energy background events, 
that are known to be preferentially located 
in the border of the detector \cita{hirata}. 
However, before proceeding we wish to stress 
the importance of a detailed quantitative 
study of the detector response to a sample of 
uniformly distributed supernova neutrino 
events (including fluctuations, realistic light 
propagation, role of the energy spectrum, 
imprecise measurement of the position, {\em etc.}) 
and we also warn the reader against the possibility of 
significant systematics effects, of which we are not aware.

\begin{figure}[t]
$$\includegraphics[width=.37\textwidth,angle=270]{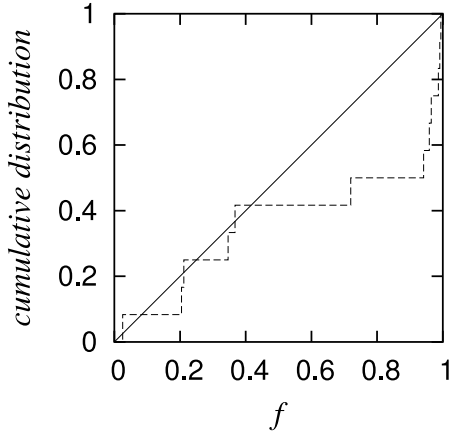}$$
\caption{\em The volumetric distribution of the 12 KII
candidate supernova neutrino events (dashed line) compared with a
uniform distribution (continuous line).\label{fig1}}
\end{figure}

As can be seen in table~\ref{tab1}, there is no particular
feature of the observable $\theta_\perp$.
Similarly for the distance $D_{out}$, which is
usually large, except for the events 10 and 2.
Anomalies emerge instead in the distributions of $D_{in}$,
of the minimal distance from the wall $d_{min}$
and of the volumetric coordinate $f$: all 
these show that there are events quite close
to the surface of the volume. In particular,
$d_{min}$ is smaller than 10 centimeters for the events 2, 3, 4, 8 and 10
and $f$ shows that the same events are contained in the outermost 4~\%
part of the 2140~ton volume, see figure~\ref{fig1}.
Performing a Smirnov-Cram\`e{}r-Von Mises 
(SCVM)~\cita{scvm_1,scvm_2} test for the hypothesis that all 12
events come from a uniform distribution 
we get a goodness of fit of 4.6~\%.
When we perform a test suited to reveal features in the border,
this hint for a deviation from a uniform distribution 
becomes stronger. In fact, the goodness of fit is 0.3~\% 
with the traditional version of the 
Anderson-Darling (AD)~\cita{scvm_1,ad} test,
whereas the one sided version 
of this test gives 0.04~\%: see App.~\ref{app:cc}
for details. Attributing the event number 6 to background
the goodness of fit becomes even lower, for the 
simple reason that this event is not located in the 
border. We can regard these results as a suggestion that
there is contamination from background 
events located close (or at) the border 
of the volume that was used for the analysis.
We interpret this as a 3 $\sigma$ hint, with
reference to the numerical value of the traditional AD test.

It is important to repeat that the assumption of a 
uniform distribution of neutrinos 
is valid whatever was the true (energy and/or time)
distribution of SN1987A neutrinos.
To the best of our knowledge, the analysis of the spatial distribution
presented here and elaborated further in 
App.~\ref{app:cc} and Sect.~\ref{discu} is original.
In \cita{hirataPRL} we read that ``the distribution of the events 
presented here is consistent with a uniform distribution'', but  
unfortunately, the procedure adopted, 
the statistical test performed and 
the significance level  are not indicated, 
so it is not possible to compare the results.

\subsection{\sf\color{verdon}  Energy distribution\label{ed}}
In this section, we compare 
the observed and expected energy distributions.
In the first part (section~\ref{pp1}) 
we assess on quantitative basis
the `problem' of the energy distribution,
in the second part (section~\ref{pp2}) 
we discuss the probabilities that the 
individual events to be due to background, that
we then use to make statements on the expected number of events and 
on significance of the observations.

\subsubsection{\sf\color{verdon}  Comparison of the expected and the measured energy 
distributions\label{pp1}}
\begin{figure}[t]
\centerline{\includegraphics[width=.37\textwidth,angle=270]{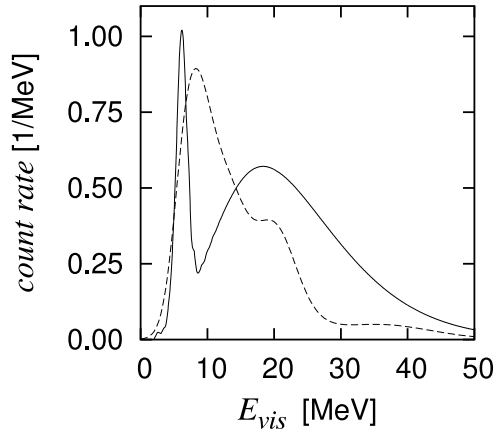}}
\caption{\em 
The continuous, bimodal curve is the 
count rate expected in KII 
from the observed background (low energy peak) and 
the supernova neutrino signal, 
see text for a description.
For visual comparison, we show also the spectrum 
obtained summing the Gaussian energy distributions 
of the 12 observed events, with $E_i$ and $\delta E_i$ 
as given in table~\ref{tab1} (dashed curve).
\label{fig2}}
\end{figure}

The expected event distribution is
the continuous, bimodal curve in figure~\ref{fig2}. 
The two distinct components are simply the 
(measured, low-energy) background and the 
(assumed, high-energy) signal due to supernova neutrinos,  
discussed in the beginning of Sect.~\ref{sep}.
Clearly, the agreement of the theoretical distribution
(the continuous curve in figure~\ref{fig2})
with the 
observations (the dashed curve in figure~\ref{fig2})
could be improved if the theoretical parameters were 
very different from what we expect, as assumed in~\cita{mira} 
and~\cita{luna}. In fact, one could be tempted to argue 
from figure~\ref{fig2} that it is {\em necessary}
to assume a significantly lower value of the 
average energy of neutrinos than assumed in equation~\ref{parama}. 
But, is it really correct to conclude 
that the apparent disagreement is not due to a fluctuation?
We performed a standard SCVM test for the null hypothesis
described in Sect.~\ref{sep} (or in figure~\ref{fig2}) 
finding that
\begin{quote}
{\em the assumed energy distribution 
(measured background + expected signal)
should not be rejected at the 
11~\% significance level.} 
\end{quote}
Thus, we believe that we are justified in assuming 
that the theoretical distribution (that was motivated previously)
is not in contradiction with KII observations.

Before proceeding, however, it is useful to 
explain better in which sense one could claim 
the existence of a `problem' with 
low energy SN1987A 
neutrinos. Suppose we ask whether the KII data could  
come from the assumed signal 
{\em setting the background to zero};
using again a SCVM test,
we find that this hypothesis 
can be accepted but only  at 0.4 \% significance level 
(2.9 $\sigma$).
In other words, we find that 
the assumed signal is in disagreement with the data
if we posit that the background is absent.
However,  in view of the preceding discussion, we believe 
that this position is hardly tenable
and, as we demonstrated here, it has an important impact 
for the interpretation of the data.
So, we assume the concomitant presence of signal and 
background. The next step is to explore what are the quantitative 
implications of this position, in particular, 
for what concerns the background.\footnote{One may ask 
why we do not proceed similarly 
for the spatial distribution. In Sect.~\ref{sd} we 
test only whether a subset of data is uniformly distributed ({\em i.e.}, 
whether the subset can be attributed to supernova neutrino events), here 
we give the energy distribution of the signal 
{\em plus} the background and compare with the observations. 
The reason of this asymmetric procedure 
is simply that we do not have the {\em a priori} information on 
the spatial distribution of the background, as 
recalled in the beginning of Sect.~\ref{sep}; we only know that 
the events with $N_{hit}\ge 23$ ``are consistent with 
higher-energy products of radioactivity 
at or outside the tank wall'' and we keep 
in mind the message of figure~5 of~\cita{hirata}.}

\subsubsection{\sf\color{verdon}  Analysis of the individual events\label{pp2}}
In \v{C}erenkov type detector it is not possible to state 
whether a single event is due to signal 
or to background. However, knowing the energy distribution of the 
signal and the one of the background,
it is possible to assign to each event
a probability $P_i$ that it resulted from background 
(and a probability $1-P_i$ to result from signal).

This can be done proceeding as follows. 
The probability 
per unit second $B_i$ (background rate) that an event 
with energy $E_i$ results from background is known
experimentally. 
The data that we use and the numerical values
of the $B_i$ are described in App.~\ref{app:aa}.
In order to obtain the signal rate, instead, we describe each event 
with a Gaussian distribution centered at the energy of
the event $E_i$ and with a width $\delta E_i$ as measured:
\begin{equation}
\rho_i(E)\propto \exp\left[{-\frac{(E-E_i)^2}{2\ \delta E_i^2}}\right]
\label{rrrr}
\end{equation}
Next, we calculate the convolution 
integral ${\cal I}_i$
of this Gaussian with the expected distribution of
the signal. We obtain the corresponding signal rate 
$ S_i={\cal I}_i/T $ simply 
dividing by the time $T$ from the 
first to the last event given in eq.~\ref{times}.
In this way we can 
compare the expectations for the background and for the signal, 
and we can calculate the adimensional quantity
\begin{equation}
P_i=\frac{B_i}{B_i+S_i}
\label{defb}
\end{equation}
that can be thought of as the probability 
that an event is due to background. The approximate values 
of $P_i$ are 
given below (0 means `small' with the given precision):
\begin{center}
\begin{tabular}{c||ccc|ccc|ccc|ccc}
event  & 1 & 2 & 3 & 4 & 5 & 6 & 7 & 8 & 9 & 10 & 11 & 12 \\
$P_i$ & 0\% & 1\% & 74\% & 29\% & 1\% & 93\% & 0\% & 0\% & 0\% & 51\% &
1\% & 43\% \\
\end{tabular}
\end{center}
This table indicates
that the events 3, 4, 6, 10, 12 have 
a non-negligible chance to be due to background,
due to their relatively low energy.

These results do not change much when we modify 
the energy distribution in a manner 
that does not contradict the theoretical expectations, {\em e.g.}, 
when we use $\alpha=2$ rather than $3$, 
or when we diminish by a $1-2$ MeV the average energy 
of supernova neutrinos. {\em E.g.},
if the average neutrino energy were 12 MeV (resp., if $\alpha=2$), 
the convolution integral 
${\cal I}_{10}$ would increase by 20~\% (resp., decrease by 10 \%),
and thus $B_{10}$ would decreases, but only by~10 \% 
(resp., increase by 5 \%). It 
is pretty much evident that the probability that a 
low energy event is due to background remains 
quite large if the signal resembles 
the one shown in figure~\ref{fig2}.

It should be noted that the number of background 
events above the threshold of 7.5 MeV, that we 
evaluate {\em a posteriori} using the selected 
model of supernova neutrino emission, 
is seven times  larger than the value 
expected {\em a priori}. In other words, even if 
postulating the presence of a pair of background events
is the best comprimise we can reach with the model, 
this is not necessarily a good compromise; this could be 
alternatively interpreted as an indication 
that the conventional model 
for neutrino emission that we adopted misses  
important features of the data. More discussion later.

Armed with these results,
we find that the probability that all events
are signal is just 0.3~\%, whereas (as obvious) it is
basically impossible to have a fluctuation of 12 events
that has these characteristics. 
We calculate on this basis a significance 
of about 10 $\sigma$ that {\em at least} one of 
these events are due to supernova neutrinos, that is 
higher than the 6.5~$\sigma$ significance 
that can be estimated from figure~10(b) of~\cita{hirata}
(using only the information on the background). 
Though this result is based on the tails 
of the energy distributions, and should not  
be fully trusted quantitatively,  
it carries a reasonable and reassuring message: 
if we include the information on the expected signal,  
the evidence for a detection of neutrinos from SN1987A 
strengthens.

The most interesting and probable cases have several background events:
\begin{center}
\begin{tabular}{c||cccccc}
\# of bkgr.~events $n$ & 1 & 2 & 3 & 4 & 5  \\
probability ${\cal P}_n$ & 6\% & 26\% & 39\% & 23\% & 5\% 
\end{tabular}
\end{center}
This is easily evaluated by constructing the polynomial
$q[x]=\Pi_{i} ( 1-P_i+x\; P_i )$ 
since, if $P_i$ is the {\em background} probability,
$1-P_i$ can be thought of as the {\em signal} probability.
The coefficient ${\cal P}_n$ of $x^n$ 
counts the probability to have exactly 
$n$ background events; thus 
we should simply expand the polynomial 
$q[x]=\sum_n {\cal P}_n x^n$.

%%%

Similar results 
are presented in table~VI of \cita{lambLoredo}. 
But in that work 
the main issue is the study of the time distribution of neutrinos, 
whereas the average energy of supernova neutrinos 
is considered a free parameter whose value is decided from 
a global fit. Here, in order to discuss whether KII 
observations necessarily indicate an 
excess of low energy neutrinos, we 
followed a different approach, and adopted a
(fixed, pre-selected) value of the $\bar{\nu}_e$ average energy that
is compatible with the calculated 
ones and with IMB observations. Other comparisons with the 
reference results obtained by Lamb and Loredo are presented in the following.

Let us repeat that the expectations on the 
background here discussed are obtained 
{\em a posteriori} adopting a pre-selected model of supernova emission. 
The {\em a priori} expectation on the background are discussed 
in the next Section, and a detailed comparison of the 
outcomes is offered in table \ref{tab2} and discussion given there.

In summary, we verified: 
(1)~that the observations do not contradict seriously the 
theoretical expectations on the energy distribution of 
supernova neutrinos; 
(2)~that this distribution, along with
the measured energy distribution of the background, 
suggests the hypothesis that 2 to 4 low energy events in 
KII dataset are due to background.

\subsection{\sf\color{verdon}  Energy threshold\label{et}}
The criterion adopted by Kamiokande-II collaboration
\cita{hirataPRL} to separate the signal from the background 
is the energy threshold of 7.5~MeV 
(corresponding to about $N_{hit}=20$)
used in solar neutrino analyses.
In fact, in the abstract of \cita{hirataPRL} we read: 
``the signal consisted of 11 electron events of energy $7.5$ to $36$ MeV'';
in the text: ``event number 6 has $N_{hit}<20$ and has been
excluded from the signal analysis''. 
In our opinion, this criterion is perfectly fine
if the aim of the analysis is to claim that there is an excess that cannot
be explained as a fluctuation of the background; it is less fine if the 
aim is to investigate the properties of the signal. 

We propose a number of critical remarks 
on the exclusive use of this criterion:\newline
{\em 1)} {\em A priori}, 
there is no warranty that a quantitative criterion that works 
well for solar neutrinos works also for supernova neutrino signal.
In fact, solar neutrino events are directional, 
whereas $\bar{\nu}_e p\to n e^+$ 
events are not, the interaction rates of solar and supernova 
neutrinos are not the same, and the energy distributions are  
also different. However, when this criterion is adopted and
{\em when a time window is selected} one can check the 
expected number of background events above threshold: from fig.~10
of \cita{hirata}, we deduce that in the time window 
spanned by the 12 events, eq.~\ref{times}, 
this makes 0.272 events. Though this is not very large, it 
is not completely safe too.
\newline
{\em 2)} In general, the choice of a criterion 
to distinguish signal and background 
is not independent from the signal we want to reveal. 
For instance, the model for supernova neutrinos 
considered here suggests that the criterion of setting the  
threshold at 7.5~MeV is insufficient
to ensure that, above, we have only supernova neutrinos:
in fact, we find that the background is smaller than 5~\%
only above 10 MeV, corresponding to about $N_{hit}=26$
(see figure~\ref{fig2}). \newline
{\em 3)} Finally, it is possible to 
argue against the procedure of excluding 
{\em only} the event number 6 
from the supernova dataset:
In fact, removing the events 6, 
the confidence level that we have 
a uniform volumetric distribution 
halves when we use the AD or the SCVM test
(see App.~\ref{app:cc} and later on).
In other words, the likelihood 
diminishes significantly {\em unless} other 
signal events are also assigned to background.
It should be clear that this last argument 
is completely independent on the 
model of supernova neutrino emission.

While we have no reservation to accept the hypothesis that
the event 6 is due to background (see App.~\ref{app:aa}), the arguments
above suggest that it could be not the only background 
event in the time window of eq.~\ref{times}.
As we discussed in the previous section, we believe that the best 
approach to extract and study 
the signal is to quantify the probability that 
an event is due to background; 
however, it is useful and instructive to consider also  
the simpler criterion of setting the energy threshold at 7.5 MeV
(the comparison of these criteria will be done in Sect.~\ref{mid}). 
More specifically, we will use two versions of this  
criterion:
\begin{enumerate}
\item[V1] In the first version, we simply discard the event number 6,
and consider the events that fall under the threshold 
within 1 $\sigma$ 
(namely, the events 3,4,10,12)
as suspect. Thus, we will attempt to attribute some of 
them to background, in order to see if the 
volume distribution improves or not. 
\item[V2] In the second version, we will do the same, except for assigning 
an additional {\em penalty factor} each time we attribute one of 
these four events to background. The penalty factor is simply
the Poisson probability for $n$ background events, where the 
average value is set to $b=0.272$ (for instance, the penalty 
factor to have 2 such events is 2.8\%).  In this way, we 
somehow describe what we know  on energy distribution of 
the background, 
though we are ignoring any {\em a priori} 
information on the energy distribution of the signal.
\end{enumerate}
These two versions of the criterion 
are used in table~\ref{tab3} of the next Section;
see also the second line of table~\ref{tab2}.

\section{\sf\color{verdon}  Average energy of SN1987A neutrinos 
accounting for background \label{discu} \label{ce}}
As we repeatedly argued, in order to address the question on 
supernova neutrinos raised in the title of the paper 
we deem it necessary to discuss the role of 
background events in KII dataset. Here we collect 
the available information for 
such a discussion and draw our conclusions on whether
there is a problem with low energy SN1987A neutrinos.
As discussed in Sect.~\ref{sep}, 
we will assume that all the events except those that are assigned to
background are due to $\bar{\nu}_e p\to n e^+$ 
events (an extension of the `inverse beta decay hypothesis').
Thus, the identification of the events due to
SN1987A neutrinos becomes equivalent to 
the identification of the background events.\footnote{Other 
approaches are in principle possible. For what concerns 
misidentification of signal against background, 
one could decide to accept only the events in the fiducial volume;
if we use the
definition of~\cita{hirataPHD} 
we are led to keep only the five events 1, 6, 9, 11, 12.
A similar possibility would be to reject all events
under a `fiducial threshold' that, according to our expectations, should
be around 10 MeV (see Sect.~\ref{et}); 
again in this way we would remove several events
including one of the previous subset.
Both procedures should provide us with relatively safe results,  
but would amount to largely diminish the
information on the characteristics of the supernova burst 
that we can extract from KII.
In view of the small dataset, 
we believe that the most useful approach is to make
an assessment on the background.}

The first task is to identify the main candidate background events,
and for this aim we use the tools of Sect.~\ref{sep}.
%% Before proceeding we remark that, 
%% with the conventional criterion of 5 \% confidence level, 
%% it is not possible to exclude that 
%% the KII dataset contains 3 or even 4 background events,
%%: the hypothesis that we have 
%% 3 or 4 background events in the interval $t\in [0,T] $
%% should be considered as reasonable. 
We begin listing again 
(a)~the events that are closer to the border,
namely: 2, 3, 4, 8, 10 and  
(b)~the events with relatively low energy:
3, 4, 6, 10, 12 with $N_{hit}$ equal to 25, 26, 16, 21, 24.
Thus, we select the events 3, 4, 6, 10, 12 for 
further discussion:
\begin{enumerate}
\item[6] This is the lowest energy event and it has
a high probability to be due to the `diffuse' 
component of the background ({\em e.g.}, internal radioactivity, 
radon, cosmic ray induced or neutrons). 
Being under the threshold of software analysis,
$N_{hit}=20$, it is removed in most 
investigations of supernova neutrinos
(see Sect.~\ref{et} and App.~\ref{app:aa}).
\item[10] Besides being a low energy event, this event
is known to travel a few tens of centimeters in the 2140 ton detector
volume and to be very close to
the lateral surface {\em and} to
the upper plane of the detector--it is in the edge of the volume.
We are not able to assign a strong quantitative significance to
this information, but we believe that it puts a second
red herring on this event.
\item[3, 4] These two events are quite similar; they both have low
energy and are both very close to the border of the 2140 ton volume
used in the analysis. Thus, rejecting one or the other
from the supernova neutrino sample produces a similar effect.
\item[12] This is just the last event, with an energy similar to
the event number 10, and thus (depending on the true 
time distribution) could fall in a region where few 
signal events are expected. 
It is located far from the border, though.
\end{enumerate}

\subsection{\sf\color{verdon}  A first global approach}

Now we attempt to attribute some of the 
events mentioned above to background, testing
whether the spatial and the energy distributions of supernova neutrinos 
improve or not. The results are shown in table~\ref{tab2},
where we give:
%% (a)~the probabilities $\pi[n]$ to have $n$ background events and
%% $12-n$ signal events, assuming that also the signal is Poissonian
%% with the average number of events
%% given by our model,  $s=11.8$ (see Sect.~\ref{nu});
(a)~the Poisson probability to have a certain number of background events 
above the threshold (see Sect.~\ref{et});
(b)~the significance levels assuming
a uniform volumetric distribution, evaluated respectively using the
SCVM and the two AD statistics described in App.~\ref{app:cc}
(see also Sect.~\ref{sd});
(c)~the probability of a given distribution of signal
and background, that in the case where all events 
are declared to be signal is $\Pi_i \ (1-P_i)=0.32$~\%, but {\em e.g.}, 
increases by the factor $P_6/(1-P_6)=14$
removing the event number 6 (see Sect.~\ref{ed});
(d)~the average energy of the signal sample,
that is the aim of our discussion.

\begin{table}[t]
\begin{center}
{\footnotesize
\begin{tabular}{|l||c|cc|cc|c|c|}
\hline
Events removed & none & 6 & 10 & 6,10 & 3,10& 3,6,10 & 3,4,6,10 \\ \hline
%(a) bkgr.\ \& sign.
%& 12\% & 28\% & 28\% & 30\% & 30\% & 19\% & 8\% \\ 
(a) bkgr. above $E_{th}$
& 76\% & 76\% & 21\% & 21\% & 3\% & 3\% & 0.3\% \\ 
(b) Vol.distr.,\ SCVM & 4.6\% & 2.4\% & 12\% & 7.2\% &26\% & 18\% & 42\% \\
(b) Vol.distr,\ AD-2side & 0.3\% & 0.1\% & 2.2\% & 1.0\% &6.9\% & 3.8\% & 21\% \\
(b) Vol.distr,\ AD-1side & .04\% & .02\% & 0.4\% & 0.2\% &1.5\% & 0.8\% & 7.8\% \\
(c) Energy distribution & 0.3\% & 4.6\% & 0.3\% & 4.7\% & 1.0\% & 14\% & 5.6\% \\ 
\hline
(d) Aver.\ energy [MeV] &
$15\pm 2.3$ &
$15\pm 2.4$ &
$15\pm 2.4$ &
$16\pm 2.5$ &
$16\pm 2.5$&
$17\pm 2.6$&
$18\pm 2.7$ \\ \hline
\end{tabular}}
\end{center}
\caption{\em Impact of various assumptions on the background events;
first row, the events assigned to background sample; 
second line, 
%the probability of a given number of background events 
%taking into account the expected number of signal events;
%third line, 
the Poisson probability of $n$
events above threshold of 7.5 {\rm MeV};
third line,
the SCVM significance level of a deviation from a uniform volumetric
distribution;  
fourth line, the same using the two sided (traditional) AD statistic;
fifth line, the same using the one sided (modified) AD statistic;
sixth line, the probability of the given energy 
configuration, evaluated with the 
considerations of section~\ref{ed};
last line, average visible 
energy of the supernova signal events.\label{tab2}}
\end{table}

\subsubsection{\sf\color{verdon}  Model dependent and independent inferences\label{mid}}
The only really model independent test 
that we are aware of is the test for 
a uniform distribution of supernova neutrinos. 
The most powerful test between the 
Smirnov-Cram\`e{}r-Von Mises and the 
Anderson-Darling test is the second one.
In fact, this is built to 
reveal deviations from the expectations 
close to the boundaries, that could mean 
the presence of background events. 
We will use this test in the following, 
alone and in combination with other ones.
We will also use the SCVM and 
compare the outcome of various procedures of analysis, 
reaching in all cases similar conclusions. 
These conclusions will be taken as a guide 
to formulate a hypothesis on the background 
and on the signal of KII dataset.

\paragraph{\sf\color{verdon}  Analysis A}
First, we present the results of the one sided (modified)
Anderson-Darling test described in App.~\ref{app:cc}. 
With the conventional 5~\% significance level, 
the only case among those of 
table~\ref{tab2} 
that should not be rejected  
is the one where all four 
candidate background events
are attributed to background
(see fifth row of  table~\ref{tab2}).
Another interesting conclusion 
that we obtain from the one sided AD test is that at least 3 events 
close to the border should be attributed to background.
The conclusions till here are independent 
on the energy distribution. When we use the 
{\em a priori} expectations on the energy distribution of the background we
conclude that the lowest energy events, among those selected by the AD test,  
have a higher chance to be due to background. 
This is the closest we can go to a model independent argument.

\paragraph{\sf\color{verdon}  Analyses B1,B2,B3}
Next, we quantify the improvements (likelihood ratios)
in the description of the supernova neutrino signal 
under the various hypotheses on the background
described in Sect.~\ref{ce}.
We will compare four different procedures.
In all of them, we perform a 
{\em traditional} (two sided)
AD test to check the uniformity of the assumed 
supernova events in the volume.
Additionally:\\ 
% B1) we check the Poisson probability for event numbers
% following Sect.~\ref{nu};\\
B1) we follow Kamiokande-II and 
assign {\em a priori} the event number 
6 to background (see Sect.~\ref{et}, first version of the
threshold criterion);\\
B2) additionally to B2, we impose a Poisson penalty 
for background events above the threshold of 7.5 MeV (see Sect.~\ref{et}, 
second version of the threshold criterion);\\
B3) we test the likelihood of the given 
energy distribution as in Sect.~\ref{ed}.\\
With any of the four procedures
we evaluate the {\em factors of improvements} 
of the given hypothesis,  choosing as a comparison the 
case when all 12 events (or 11 events in the second case) 
are considered to be due to supernova 
neutrinos.
{\em E.g.}, the procedure B3 suggests that the hypothesis that 
only the event 6 is due to background is more probable than 
the hypothesis that all events are signal, by a
factor of improvement equal to
\begin{equation}
\frac
{P_{vol.}(\mbox{all events but }6)\times P_{en.}(\mbox{all events but }6)}
{P_{vol.}(\mbox{all events})\times P_{en.}(\mbox{all events})}
=5.7
\end{equation}
(approximate values of $P$ are given in table~\ref{tab2}).
The results of these calculations are given in table~\ref{tab3}.

The procedure B1 makes very little use of the background and
of the signal; the procedure B2 uses the {\em a priori} 
knowledge on the background, but 
neglects any information on the signal; the procedure B3
uses all we know, though its reliability depends also 
on the model for supernova neutrinos we assume. 
It is difficult 
to make firm conclusions from this table, but we note that even in 
the more pessimistic procedure (B2) the presence of several
background events cannot be firmly excluded; this is because 
an {\em a priori} unlikely fluctuation of the number of 
background  events above threshold is 
compensated by the fact that some of these events 
lie close to the border of the 
detector. 
On the other hand, we see that with the procedure of analysis B3,
the only one that makes a  full use of the knowledge on the background 
and on the signal,  the factors of improvements are about 
a thousand times when all 4 candidate background events
are assumed to be background, and half of that value 
when we assign events 3, 6, 10 (or 4, 6, 10) to background.
In short, table~\ref{tab3} suggest that a compromise 
between the extreme assumptions---namely, of considering 
completely unknown or reasonably known 
the energy distribution of the signal--would admit the 
presence of several background events.
It is important to note that the results of 
the more aggressive procedure of analysis (B3) 
do not change much when we modify 
the energy distribution in a manner 
that does not contradict the theoretical 
expectations, see Sect.~\ref{ed}.
In other words, one could be tempted to argue that 
the model dependence of our stronger analysis is not large, if  
the astrophysical parameters lie in the range 
suggested by the theory and by the IMB observations: 
see again Sect.~\ref{sep} and \cita{aldo}.

\begin{table}[t]
\begin{center}
{\footnotesize
\begin{tabular}{|c||c|c|c|c|c|c||c|}
\hline
AD-2side + &  6 & 10 & 6,10 & 3,10 & 3,6,10 & 3,4,6,10 &  remarks  
 \\ 
\hline
%bkgr. \& sign. &  0.9 & 16 & 8.2 & 54 & 20 & 47 & almost model indep.\\
Sect.~\ref{et}, V1
&  $\equiv 1$ & $-$ & 8.2 & $-$ & 30 & 160 & model indep.\\
Sect.~\ref{et} V2 
&  $\equiv 1$ & $-$ & 2.2 & $-$ & 1.1 & 0.5 & model indep. \\ 
Energy distrib.   &  5.7 & 7.0 & 48 & 64 & 500 & 1100 & model dependent
\\ \hline
\end{tabular}}
\end{center}
\caption{\em Factors of improvement in the description of the 
Kamiokande-II supernova neutrino
events for the selected hypotheses on the background.
The subset of events that are assigned to background is 
in first line. In addition to the Anderson-Darling 
test for a uniform volumetric
distribution, we use the information listed 
in the first column. When we 
impose the Kamiokande-II threshold (second and third rows) the 
event 6 is discarded {\em a priori}
from the supernova dataset. Last column, comments
on the selected procedure. 
\label{tab3}}
\end{table}

\paragraph{\sf\color{verdon}  Analyses C1,C2}
Finally, one may ask what happens when 
we apply the SCVM test for  
the spatial distribution of the events, instead 
than the AD test that we used so far,
or when we change the procedure of analysis.
We show that the conclusion is 
very similar to the previous ones,
though it is true that the weight of the energy 
distribution becomes comparably more important with 
the SCVM test.

Let us begin with a description of the procedure C1.
According to our hypothesis,
any of the 12 events can be due either to background  or 
to inverse beta decay signal. Thus we consider the $2^{12}=4096$
individual possibilities, 
and for any of them we test the likelihood
of the given spatial distribution, 
using this time the SCVM test and the likelihood 
of the given energy configuration as we did in Sect.~\ref{ed} 
(we checked that with a SCVM test we obtain similar outcomes).
Next, we consider the product of the two likelihoods 
and normalize the sum of the likelihoods of the 4096 
mutually exclusive cases to unity.
We find that the occurrence of 
several background events is likely:
\begin{center}
\begin{tabular}{c||ccccccc}
\# of bkgr.~events $n$ & 0 & 1 & 2 & 3 & 4 & 5 & 6  \\
probability            & 0.1\% & 2.2\% & 14.6\% & 37.3\% & 33.8\% & 11.4\% &
0.5\%  
\end{tabular}
\end{center}
The individual cases with probability above 5 \% are the following ones:
\begin{center}
\begin{tabular}{c||c|c|c|c|c|c|}
background~events             &
3,6,10 & 
3,4,6,10 & 
3,4,6,10,12 &
3,6,10,12 & 
3,4,6 & 
3,6\\
probability ${\cal P}_n$ & 19.1\% & 18.3\% & 9.9\% & 8.1 \% & 7.7 \% &
6.8 \%
\end{tabular}
\end{center}
We note that the event number 12 is occasionally assigned 
to background because of its low energy.
What about the event number 6? 
In all those cases that have probability larger than 3 \%, this 
event is due to background. We also remark that 
the 2048 cases when this event is due to background 
have a total probability of 90\%.

\begin{figure}[t]
\centerline{\includegraphics[width=.31\textwidth,angle=270]{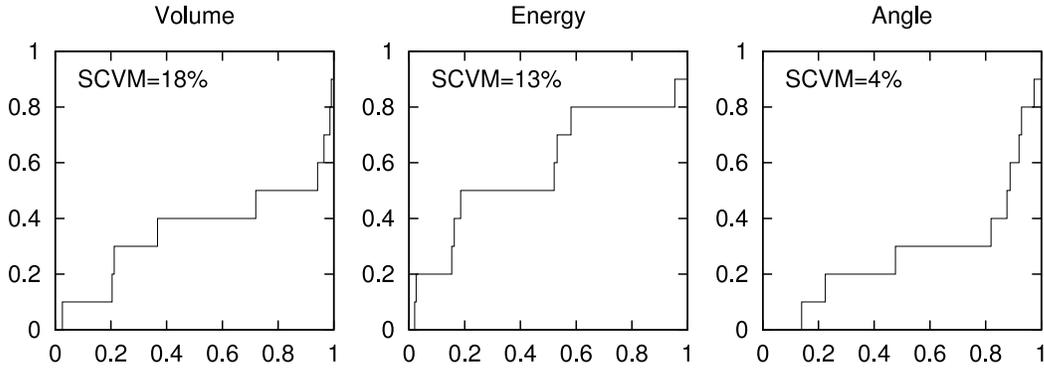}}
\caption{\em Cumulative distributions for the case when the events 
number 3, 4, 6 are assigned to background, and significance  
of the SCVM test. In the abscissa the value $F_j$ of the 
cumulative distribution (see App.~\ref{app:cc}).
\label{fig3}}
\end{figure}

In the procedure of analysis C2
we test {\em also} the SCVM likelihood of 
the angular distribution following~\cita{aldo}.
The events that are most 
commonly assigned to background are the usual ones, namely 
those figured out in the beginning of Sect.~\ref{ce}.
The individual cases with probability above 5 \% 
(the most probable individual cases) are those 
when the background is 
given by the events number 3,4,6 (15.4~\%), 3,6 (14.0~\%), 
3,6,10 (11.9~\%);  3,4,6,10 (9.1 \%), where  
in brackets we give again their {\em a posteriori} probabilities,
that by definition are normalized to give 1 when we sum 
over the 4096 possible cases.
The cumulative distributions in the volume, in the energy and in angle
for the most probable single case (the first one listed just above) 
are shown in figure~\ref{fig3}, along with the likelihood of the 
three experimental distributions.
Remarks on the persistent discrepancy of the angular 
distribution will be offered in Sect.~\ref{ddu}. 
The overall probabilities to have 2, 3 or 4 background 
events becomes 28\%, 41\% and 20\% and the probability that
all events are due to signal is~0.5\%.

\subsubsection{\sf\color{verdon}  A tentative explanation of low energy events\label{ansa}}
In view of the arguments above discussed, we formulate our hypothesis 
on the signal and on the background: 
\begin{quote}
{\em 9 (or less) among the 12 events 
observed in Kamiokande-II were due to  
SN1987A neutrinos, whereas the remaining 
3 (or more) low energy events 
were due to background.}
\end{quote}
The motivations and the merits of this position 
are discussed in Sect.~\ref{mid}.

Quite importantly,
this hypothesis allows us to argue that there is 
no excess of low energy supernova neutrinos.
Consider as a quantitative indicator 
the average visible energy of 
supernova neutrino events, 
that being weighted with the detection efficiency and the cross section
is significantly higher of the 
average antineutrino energy--see, {\em e.g.},~\cita{aldo}.
With the signal described in the beginning of
Sect~\ref{sep},  the 
average visible energy of supernova neutrino events is 
expected to be 
\begin{equation}
E_{vis}=20.6\mbox{ MeV}.
\end{equation}
This can be compared with the value given from the data, 
using the formula 
\begin{equation}
\langle E\rangle\pm \sqrt{\frac{\langle E^2\rangle -\langle E\rangle^2}{n} },\
\ \ \ \langle E^a\rangle \equiv \sum_{i=1}^{n} \frac{E_i^a}{n}
\end{equation}
where $n$ is the number of events due to supernova neutrinos.
If we assume that all events are 
due to signal ($n=12$), we find that an average energy  
that is 2.6 $\sigma$ lower than the expectations. This 
discrepancy is very similar to the 2.9 $\sigma$ `problem' of  
low energy neutrinos that was formulated in the 
beginning of Sect.~\ref{ed}: one can say that  
the selected observables catches the crucial 
feature of the `problem'.
Next we deduce the value of the
average visible energy of supernova neutrino events from the data,  
in the hypothesis on the signal formulated just above.
In the case when we assign 
the events 6, 10, 3 (resp., 6, 10, 3, 4) 
to the background  the average visible energy of supernova neutrino events 
is 1.4 (resp., 0.9) standard deviations 
below the expected value (approximate values are
in the last line of table~\ref{tab2}). In other words, 
they are in reasonable agreement.
Thus, if 3 or 4 low energy events were caused by background
the question raised in the title of the paper 
should receive a negative answer.

\subsection{\sf\color{verdon}  A second global approach\label{2ga}}

The previous interpretation leaves something to be desired. 
In fact, we expect only 0.272 background 
events above the threshold of 7.5 MeV 
on average in the time between the first and the last event, so  
there was only a 3~\% probability {\em a priori} 
that  2 or more such background events occurred
(see third row of table~\ref{tab2}). 
In other words, even if the invoked background fluctuation 
cannot be just dismissed, it is significantly larger 
than {\em a priori} expected.
Furthermore, the event number 10 is in a region 
of the detector much noisier than on 
average (compare table \ref{tab1}, Sect.~\ref{ce} 
and fig.~5 of \cita{hirata}) but
the events 3 and 4 seem to be located in a relatively
safer region. In other words, the hypothesis discussed previously 
is the {\em best} one within the considered 
theoretical context, but it is not particularly {\em good}:
this can be seen quite clearly from figure~\ref{fig2},
that shows that the low energy events we are discussing 
are just in between the signal and the background 
distributions.
So, we are motivated to consider whether we can 
evade the conclusion we reached and/or how we can 
reach a more satisfactory conclusion. 
We will show  that in order to address 
at least partially these needs 
it is sufficient to consider a more accurate model for the 
supernova neutrinos, in particular, for what regards the time 
evolution of the neutrino luminosity. This possibility is 
quite appealing, since it is largely within 
the conventional expectations.

\subsubsection{\sf\color{verdon}  A more refined hypothesis 
on neutrino luminosity\label{accrocc}} 
A phase of intense neutrino luminosity is  
expected during the first hundred of milliseconds from the
collapse. This phase has a marked non-thermal  character \cita{nad}
and it is thought of as to play a key role for the explosion 
\cita{del1,del2}, \cita{del3}.
In the following we term this hypothetical phase 
as ``accretion'', and the subsequent thermal
phase as ``cooling''. The latter phase is the one 
considered till now. For the cooling phase, 
we will continue to assume equipartition,
the same amount of emitted energy, the same average energy as in 
eq.~\ref{parama}, and an emission time of the order 
of the time from the first to the last event ($T$ in eq.~\ref{times}).
Accretion instead lasts just a fraction of second
and is much more luminous. We further assume that
during accretion equipartition is violated by an excess of 
electronic neutrinos and antineutrinos: to be definite, 
we suppose that, on average, there are 3 $\nu_e$ 
and 2 $\bar\nu_e$ each $\nu_\mu,\nu_\tau,\bar\nu_\mu,$ or $\bar\nu_\tau$.
We describe the $\bar{\nu}_e$ luminosity in 
the simplest manner we can conceive:
\begin{equation}
{\cal L}(t)=\left\{
\begin{array}{lclr}
\displaystyle \frac{{\cal E}_a}{\tau_a} & \mbox{for }t<\tau_a  &
\mbox{\small (``accretion'' or non-thermal phase)}\\[2ex]
\displaystyle \frac{{\cal E}_c}{\tau_c} \times e^{\frac{\tau_a-t}{\tau_c}}& 
\mbox{for }t>\tau_a 
& \mbox{\small (``cooling'' or thermal phase)}\end{array}
\right.
\end{equation}
where we select reasonable values  for the 
parameters, namely 
${\cal E}_a=8\times 10^{51}$ erg and $\tau_a=0.5$ s for accretion, 
and ${\cal E}_c=4\times 10^{52}$ erg and $\tau_c=T$ for cooling. 
{}With these numerical values, we can derive several other
interesting quantities: 
1)~the total emitted energy 
${\cal E}_b=4.5 {\cal E}_a+6 {\cal E}_c=2.76\times 10^{53}$ erg;
2)~the fraction emitted during accretion 
$4.5 {\cal E}_a/{\cal E}_b=13$~\%; 
3)~the electronic lepton number emitted 
during accretion, that corresponds to about 
half solar mass of iron if the average energy 
of $\nu_e$ is of 12 MeV;
4)~the energy of 20 times $10^{51}$ erg (=20 foe=20 bethe)
emitted in $\nu_e$ and $\bar{\nu}_e$, 
that could be sufficient for the explosion.
None of these numbers should be taken too literally,
but they are comfortably within 
the present theoretical expectations.

\begin{figure}[t]
\centerline{\includegraphics[width=.37\textwidth,angle=270]{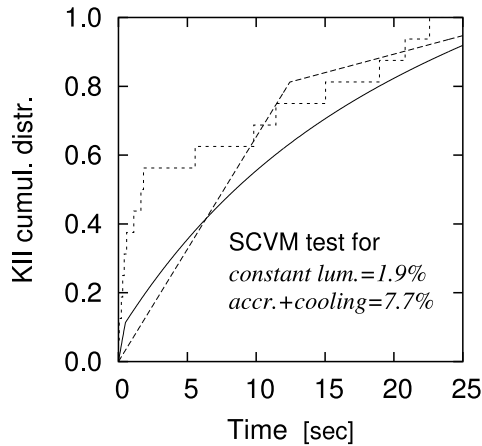}}
\caption{\em Cumulative time distribution for the 16 events seen
by KII in a time window of 30 seconds,
compared with the expected distributions for an 
emission with constant luminosity 
(long dashed) and 
with a `standard' accretion phase (continuous line).  
\label{fig4}}
\end{figure}

Let us come now to the interpretation of the events.  
Since the average energy is supposed to stay unchanged 
and the total emitted energy is almost the same,  
the total number of supernova events and their
energy distribution do not change much: 
only their time distribution changes. 
Before explaining why this position    
{\em is} relevant to the interpretation of the low energy KII events, 
we begin noticing that the expected rate of event accumulation,
\begin{equation}
N(<t)=0.187\ t + \left\{
\begin{array}{ll}
\displaystyle 2.0\ \frac{t}{\tau_a}  & \mbox{for }t<\tau_a  \\[2ex]
\displaystyle 2.0 + 11.8\ \left( 1-\exp\left[\frac{\tau_a-t}{\tau_c}\right]\right) & 
\mbox{for }t>\tau_a 
\end{array}
\right.
\end{equation}
compares reasonably well with the KII data, 
since the (SCVM) GOF is 7.7\%. In the previous equation
we show the contributions of the background--first term--and of the 
signal separately; the number of events during cooling 
is the same as before, 11.8.
For comparison, the hypothesis  
that the signal accumulation rate is
$11.8/T$ for $t<T$--that is, supernova neutrinos are emitted with a
constant luminosity in an interval of $T$ seconds--has 
a GOF of 1.9\% only. 
See figure \ref{fig4}, where we show the sequence of events 
observed by Kamiokande-II in a time window of 30 seconds.
The last events are well explained by the expected 
background curve; 
instead, the remarkable feature of data close to 
$t=0$ is only partially explained by our assumption that a  
standard accretion phase of neutrino emission also 
occurred. Indeed, the expected number of events 
during accretion is only 2.0 (a bit less than the value 
$2.4 =11.8 \times  {\cal E}_a/{\cal E}_c$ due to oscillations).
If it was possible to double the number of events during accretion, 
the GOF would increase to 30\%. The other parameters of emission 
play a less crucial role for the subsequent discussion.\footnote{It 
is possible to somewhat 
improve the fit to the data using slighly different model parameters 
(e.g., a value of $\tau_c$ some times smaller than we assumed) but 
this would not change radically our conclusions.}
Similarly, if we would arbitrarily omit the event number 3
and the  event number 6 from the dataset, the GOF 
would increase to 25\%. For other recent analyses of time distribution
that also find an evidence of an initial luminous phase see
\cita{am,lambLoredo,dr}.
Now, we pass to discuss the impact of the updated 
hypothesis on the interpretation of the events.

\subsubsection{\sf\color{verdon}  Accretion and the meaning of the early  events\label{lop}} 
The theoretical context defined in section \ref{accrocc} suggests 
strongly the possibility that some of the early events seen in KII 
occurred during a phase much more luminous than average:
${\cal E}_a/\tau_a \sim 5\times  {\cal E}_c/\tau_c$.
Identifying, arbitrarily but not unreasonably, the 
arrival of the first neutrino to the Earth with the time of 
first detected event, this possibility concerns 
the events number 1,2,3,4. Thus, the analysis  of section~\ref{ed}
should be updated, taking into account that the signal rate
in the denominator depends explicitly from the time $t_i$
at which the 
individual event occurred. More precisely, the factor $S_i$
in equation \ref{defb} should be multiplied by $\xi(t_i)$
that depends on the instantaneous luminosity: 
\begin{equation}
{\xi}(t)=\left\{
\begin{array}{ll}
\displaystyle 4.15 & \mbox{for }t<\tau_a  \\[2ex]
\displaystyle \exp\left[ {\frac{\tau_a-t}{\tau_c}} \right]& 
\mbox{for }t>\tau_a 
\end{array}
\right.
\end{equation}
Of course, the value of
${\cal E}_a/\tau_a$ (the luminosity during accretion) 
determines the numerical factor 4.15, and the value of the 
luminosity in the beginning of the cooling phase ${\cal E}_c/\tau_c$ 
gives $\xi(t)=1$ just after $t=\tau_a$. A straightforward 
calculation gives the new values for the probability $P_i$ 
that an individual event is due to background:
\begin{center}
\begin{tabular}{c||ccc|ccc|ccc|ccc}
event  & 1 & 2 & 3 & 4 & 5 & 6 & 7 & 8 & 9 & 10 & 11 & 12 \\
$P_i$ & 0\% & 0\% & 41\% & 9\% & 1\% & 94\% & 0\% & 0\% & 0\% & 67\% &
2\% & 67\% \\
\end{tabular}
\end{center}
(compare with Sect.~\ref{pp2}).
The most striking consequences 
of the new hypothesis is the increased probability that  
the event number 4 is due to supernova signal, whereas
the late events have a higher chance to be caused 
by background. Due to
these features, the expectations on the number 
of background events, based on the energy distribution, are not 
radically modified:
\begin{center}
\begin{tabular}{c||cccccc}
\# of bkgr.~events $n$ & 1 & 2 & 3 & 4 & 5  \\
probability ${\cal P}_n$ & 7\% & 28\% & 41\% & 20\% & 2\% 
\end{tabular}
\end{center}
The probability that all events are signal is 0.36\%,
and we can exclude at 10 $\sigma$ 
that all events are due to background. However, there is 
an important difference with what we found based on the 
results of section \ref{ed}:
The events 3,4 (10) that are located close to the border of the detector
have a smaller (larger) chance 
to be due to background, and this has an impact on global 
analyses. Using as a reference analysis C2 described above,
we find the following results, that can be compared with those 
in section~\ref{mid}: 
\begin{itemize}
\item The probability that no background events 
occurred becomes 0.9\%. This is small is absolute 
terms, but larger than estimated above 
using only the energy distribution.
\item The probability that a certain number of background 
events occurred is:
\begin{center}
\begin{tabular}{c||cccccc}
\# of bkgr.~events $n$ & 1 & 2 & 3 & 4 & 5  \\
probability ${\cal P}_n$ & 11\% & 36\% & 36\% & 14\% & 2\% 
\end{tabular}
\end{center}
that is rather similar to the values calculated just 
above, based on the expected energy distribution only.
\item The individual cases that occurred with 
probability above 5\% are:
\begin{center}
\begin{tabular}{c||c|c|c|c|c|c }
background~events             &
6,10 & 
3,6,10 & 
6,10,12 &
6& 
3,6 & 
3,6,10,12\\
probability ${\cal P}_n$ & 
18.9\% & 
12.0\% & 
9.5\% & 
7.9 \% & 
7.0 \% &
5.7 \% \\ \hline
volumetric SCVM  & 7.2\% & 18.0\% & 3.3\% & 2.4 \% & 6.5 \% & 9.9 \% \\
angular SCVM & 3.3\% & 1.2\%  & 1.8\% & 8.4 \% & 4.0 \% & 5.2 \% 
\end{tabular}
\end{center}
where we give  
also the GOF of the volume 
and of the angular distributions in the last two lines.
Note that these 6 cases alone cover 61\% of the 4096 possibilities.
\item The event number 6 is due to background in 90\% of the cases.
\end{itemize}
In summary, the existence of a accretion phase brings 
a new twist in the analysis of the backgrounds, and leads to somewhat 
different conclusions especially regarding the interpretation of 
the event number 4. However, the new analysis still leads us to expect
that some low energy events among the events number
3,6,10,12 could be attributed to background.

\subsubsection{\sf\color{verdon}  Again on the interpretation of low energy events\label{ansa2}}
In view of the discussion of the previous section, 
the interpretation proposed in section~\ref{ansa} 
does not need a real revision, except, possibly, replacing  
the event 4 with the event 12 in the role of candidate background event.

However, the wider theoretical context also allows us to take 
a new point of view, or in other words, it is possible to formulate 
another way out from the `problem' of low energy events. In fact,
one could argue that the events occurred during the non-thermal phase
(accretion) should not be compared with the ones occurred later.
In particular, the average energy of the first four KII events 
is $12.6\pm 2.4$~MeV, that is pretty much lower than the one of the 
next seven events (the event number 6 being attributed to 
background), namely $17.1\pm 3.3$~MeV. 
Of course this is due to the presence of a few low energy events 
among the first events (the events 3 and 4).
But while in section~\ref{ansa} we regarded them as 
as possible background events, here we are lead to suggest 
that they (or at least some of them) 
are due to a {\em peculiar} phase of 
neutrino emission, that in principle could be characterized by
a lower average energy than during thermal phase. 
In short, the fact that KII measured several low energy events 
can be at least {\em partly} attributed to the fact 
that some of these events occurred during accretion.
In order to have an acceptable volumetric distribution,
it is sufficient to assign only one additional event 
between 3 and 10 (along with event 6)
to background.

\section{\sf\color{verdon}  Discussion\label{ddu}}
We have shown in Sect.~\ref{discu} that the 
agreement between the observed and expected 
average energy of supernova neutrinos improves 
significantly assuming that 
2-4 of the low energy events in KII 
are due to background, and that this interpretation is supported 
by the volumetric distribution of the events. A few of these 
low energy events could also belong to a peculiar 
phase of neutrino emission (``accretion''), 
 contributing to solve 
the `problem'.  In other words,  we believe that  
the null hypothesis ({\em i.e.}, the minimal model)
is not significantly challenged by the excess of 
low energy events observed by Kamiokande-II 
(or: we have enough freedom 
to evade the conclusion that there is a problem with the 
conventional expectations). 
Here, we would like to offer various remarks on this conclusion.

%%%

\paragraph{\sf\color{verdon}  New tools used.}
A new crucial information that permitted us to reach new conclusions 
is the analysis of the spatial distribution of the events (Sect.~\ref{sd}).
The validity of the hypothesis on the background that we 
formulated in Sects.~\ref{ansa} and \ref{ansa2}
rests also, to a certain extent, on the assumption we made 
on the signal of supernova neutrinos
described in the beginning of Sects.~\ref{sep} and \ref{accrocc}.
In our view, these should be considered a conservative 
and reasonable assumptions on the signal. They are consistent with 
the IMB observations that are basically 
background free. Furthermore, the conclusions we reach remain 
valid under small modifications of the model for supernova neutrinos,
and do not contradict (but rather strengthen) what we obtain following 
Kamiokande-II and setting the energy cut at 7.5~MeV
(Sect.~\ref{et}).

\paragraph{\sf\color{verdon}  Peculiarity of the approach.}
The simple but crucial 
observable that we have used to draw our conclusions 
is the average value of the visible energy, 
see in particular Sects.~\ref{ansa} and~\ref{ansa2}. 
A detailed comparison of the energy distributions 
in figure~\ref{fig2} would suggest two anomalies, 
namely: a)~a defect of high energy events around $20-40$ MeV and 
b)~an excess of events around 10 MeV.
While we argued that the low energy anomaly  
could be due to background events (or to the fact that 
several events occurred during accretion) we are unable to 
explain the high energy anomaly if not resorting to 
a fluctuation of the data. However, `global' observables 
are safer against the effects of the fluctuations;
thus, in view of the low number of collected events
we emphasized the comparison of average visible energies
rather than the comparison of energy distributions.
In this respect, our formulation of the `problem' 
of low energy events, or equivalently our approach to KII dataset, are 
more conservative and distinct from the ones of other 
recent analyses such as \cita{mira} and~\cita{luna}, 
and closer to the one of~\cita{aldo} (see figure~1 there).
A quantitative statement is in the 
first paragraph of Sect.~\ref{pp1}.

\paragraph{\sf\color{verdon}  Leftover possibilities} 
and/or alternative (but not exclusive) 
keys of interpretation.
It is not impossible that other reasonable
effects contribute to explain the apparent excess of low energy events
(and/or defect of high energy events) in KII dataset, for instance:\newline
$(i)$ The measured energies should not be 
thought of as Gaussians (as suggested by the reported data).
If we use errors that scale as $\sqrt{E}$ \cita{raffs}
rather than 
being constant as assumed in the rest of the paper (equation 
\ref{rrrr})
we see a modest improvement of the agreement in 
the region dominated by the background ({\em i.e.}, low energies).\newline  
$(ii)$ There is one event 
due to elastic scattering that degrades the visible energy
as allowed by the selected theoretical model of emission~\cita{aldo}. 
This option is interesting since when assigning 
events to background, the angular distribution does {\em not} 
improve much (see {\em e.g.}, figure~\ref{fig3}).\newline
$(iii)$ The neutrino flux in the thermal phase 
deviates slightly from expectations, say,
in the direction suggested by~\cita{mira} but possibly
not that radically (in our evaluations, 
the change is not large--see 
Sect.~\ref{mid}).\newline
$(iv)$ Some observed event is really due to supernova
neutrinos, but it suffers of a very poor energy reconstruction
(though, we do not have the necessary information to 
elaborate on this possibility).\newline
$(v)$ Some observed events are due to 
supernova neutrinos, not necessarily $\bar\nu_e$, 
interacting with the walls of the detector (a similar 
possibility was originally noticed 
for non-standard collapses \cita{ir}, 
but was shown to yield a relevant amount of events 
in existing detectors also for a standard collapse \cita{marco}).\newline 
$(vi)$ The time structure of the signal is significantly different from 
what we know, e.g., the phase of accretion is much more 
luminous (see Sect.~\ref{accrocc}).\newline
While most possibilities are largely within conventional expectations,
some of them go beyond the minimal model that we formulated.
This is true for the last two possibilities listed just above, 
that are however tightly connected with the most 
unknown aspects regarding the explosion, and 
(if proven to be viable) could offer us new insights
on what happened during SN1987A neutrino burst.

\paragraph{\sf\color{verdon}  Possible future developments.}
With more information, more detailed analyses of the
KII data could be possible. In fact, one should analyze 
the energy and the spatial distribution at the same time,
assuming that the observations are due to {\em several phenomena}; 
a signal of high energy
due to neutrino interactions with free protons, 
with electrons and with oxygen nuclei, 
uniformly distributed in the detector
(and possibly, also 
a hypothetical component of the signal concentrated 
on the wall of the detector);
a low energy background component, 
similarly distributed, with (in principle) 
known intensity; a more energetic background component, concentrated in the 
wall of the detector, also known (again, in principle). 
More discussion of these needs is in App.~\ref{app:aa} and App.~\ref{app:bb}.
Another interesting thing would be the calculation of 
the best fit for the supernova neutrinos 
parameters, keeping into account the presence of 
background in KII, with peculiar (and in principle known)
spatial and energy distributions. 
For lack of information--as discussed in the beginning of Sect.~\ref{sep},
compare also with the discussion of equation~3.22 in \cita{lambLoredo}--we 
have been forced to analyze the energy and the spatial distributions
separately, simply distinguishing in each analysis 
between a signal and a background component.
Perhaps, more detailed analyses like these will be carried out
in the future; in the meantime and view of what we learned here, 
we are lead to expect  as a plausible outcome 
that the event 6 should be attributed to background 
because of its low energy, and some events among 3, 4, 10 (resp., 
10, 12)  should be attributed to background mostly because of their 
spatial distribution (resp., of their relatively late 
occurrence).
%To summarize, the hypothesis that the SN1987A dataset consists of 
%8-10 events (the residual low energy events being due to background)
%makes less problematic the interpretation of the 12 KII events. 

We wish to close recalling the warning of E N Alexeev: 
``it is possible that some of the low energy events 
carry an important message, still to be understood''. 
We believe that these are wise words and we subscribe
them; as we saw, the message could be the 
existence of a luminous  phase of neutrino emission (Sect.~\ref{2ga}). 
However, we also believe that it is important to check how 
close (or how far) we can go to describe the data 
using only conventional hypotheses on supernova 
neutrinos, and this is what we tried to do in this paper. 
Certainly, the handful of events collected from SN1987A was enough
for the first observation of supernova neutrinos
(the quantitative statement on the significance of 
KII observation is in Sect.~\ref{pp2}), 
but as we showed with our analysis,
the possibility that some of these 
events are not due to $\bar{\nu}_e\ p\to n\ e^+$ 
should suggest caution whenever we attempt to infer 
the characteristics of SN1987A neutrino emission from 
the observations.

\subsection*{\sf\color{verdon}  Acknowledgments}
We thank for discussions 
F~Burgio, 
G L~Fogli,
W~Fulgione, 
P L~Ghia,
P~Lipari,
E~Lisi,
A~Mal'gin,
D~K~Nadyozhin,
O~G~Ryazhskaya,
M~Selvi,
A Yu~Smirnov,
A~Strumia,
F~L~Villante and the
anonymous Referee who offered us the occasion 
to clarify and elaborate the results of this work.

\appendix

\section{\sf\color{verdon}  The assumed background\label{app:aa}}
We interpret the curve $B(E)$ presented in fig.~2 of \cita{lambLoredo} as the 
{\em measured} background rate for the estimated event energy. 
We tested this interpretation by constructing a histogram 
from fig.~4 in \cita{hirata} and using a linear conversion factor 
between $N_{hit}$ and $E$, fixed by the correspondence 
between known values: $N_{hit}=20\Leftrightarrow 7.5$~MeV and/or 
$N_{hit}=26\Leftrightarrow 10$~MeV. Within limited statistics, the 
histogram coincides with the curve given by Lamb and Loredo (LL). Note 
that our interpretation agrees with the one of LL \cita{lambLoredo},  
who state ``KII and Baksan provided us with 
measurements of $B(E)$'' and later describe their 
fig.~2 as ``background rate measurements''.
\begin{table}[t]
{\footnotesize
\begin{center}
\begin{tabular}{c|cccccccccccccccc}
$i$&  1&2&3&4&5&6&7&8&9&10&11&12&13&14&15&16 \\  
\hline \\[-2ex]
$B_i^{LL}$& 1.6$_5$&1.9$_3$&2.9$_2$&1.2$_2$&2.1$_3$&3.7$_2$&4.5$_5$&8.2$_5$&1.5$_5$&1.5$_2$&1.9$_3$&1.6$_2$ &3.8$_2$&2.9$_2$&2.8$_2$&3.8$_2$ \\ 
$B_i$& 1.0$_5$&5.4$_4$&3.1$_2$&8.5$_3$&5.3$_4$&7.1$_2$&5.0$_6$&1.0$_5$&1.0$_5$&1.8$_2$&4.0$_4$&1.4$_2$ & 7.3$_2$&5.2$_2$&1.8$_2$&7.3$_2$ 
% \\ \hline 
%\\[-2ex]
%$E_i$& 20.0&13.5&7.5&9.2&12.8&6.3&35.4&21.0&19.8&8.6&13.0&8.9
\end{tabular}
\end{center}}
\caption{\em \label{tabloid}
$1^{st}$ line, the progressive event number; next,
the estimated background rate in $\rm {Hz/MeV}$
according to the procedure of the `convolution integrals' followed by 
LL ($2^{nd}$ line) and the procedure we
describe in the text ($3^{rd}$ line). 
%$4^{th}$ line, the energy of the event in {\rm MeV}. 
A subscript 
indicates the exponent: e.g., $3_2$ means $3\cdot 10^{-2}$.
\label{tabb}}
\end{table}
Now, since the measured background already includes 
the effects of the fluctuations in the energy measurements
(`smearing'), we estimate the background rate 
for the individual event using the central value of 
the measured energy: $B_i=B(E_i)$, see table~\ref{tabb}.  
A different prescription was followed in~\cita{lambLoredo},
namely, $B_i^{LL}$ was evaluated by taking 
the convolution integral of the curve $B(E)$ with 
the Gaussian energy distribution of the individual 
events; the values from table III of~\cita{lambLoredo} 
are given for comparison.
In our understanding, such a prescription leads to double counting 
the effect of `smearing', 
and thus (1)~to underestimate the $B_i$ for the 
events with energy close to the background peak, and 
(2)~to overestimate those in the tails of the 
$B(E)$ distribution.\footnote{We believe 
that the procedure of taking 
the convolution integrals  should be applied 
to the {\em true} energy distribution of the background, 
namely, the distribution that does not include the 
fluctuations in the energy measurement.
The convolution integral would then amount to include 
the effect of the fluctuations on 
individual basis, and could lead to a more accurate 
background evaluation.} 
The new values of $B_i$ 
have been obtained as follows: 
below 10 MeV, we obtain accurate data from the 
plot of LL, whereas, above, this is not of much use;
above 9 MeV, however, we  can use the background curves 
denoted as STEP1 and STEP2 shown in fig.3 of \cita{moriond} 
for the purpose of solar neutrino studies, 
with the conversion between energy and $N_{hit}$ given there
(STEP1 agrees with the curve of LL, where comparable).
Below 7 MeV, the new procedure leads to 
higher $B_i$ (ev.s\ 6, 13-16); 
from 7.5 to about 9 MeV, there is no 
significant difference with the procedure of LL (ev.s\ 3, 10, 12);
between 9 ad 17 MeV, the background curves in \cita{moriond} 
suggest a lower $B_i$ (ev.s\ 2,4,5,11); above--in the 
most uncertain but safest from the background point of view,
and also less crucial region for parameter 
estimation--we assume a lower background rate than 
assumed by LL (ev.s 1, 7-9). 
More precise values could be provided by the experimental collaboration
using $B(E,\vec{x})$--not the average value $B(E)$--thus 
describing the role of the position of the individual events
(similarly for detection efficiencies).
It should be noted that the changes in the values of the 
$B_i$ that we propose  
are not crucial for the analysis; their main effect is just to 
increase the {\em a priori}
probability that the event number 6 is due to background,
as in the first and most popular interpretation \cita{hirataPRL}.

\section{\sf\color{verdon}  The assumption of uniform volumetric distribution\label{app:bb}}
We analyze critically the hypothesis that the events from the
signal are uniformly distributed. 
In the relevant publications 
\cita{hirataPRL,hirata,hirataPHD,moriond} 
we were not able to find warnings 
against important systematics of this type, but we 
tentatively identified three such 
physical effects that could affect the 
observed signal events:\\[1ex]
{\em Geometry:} The events occurring close to the border 
and propagating toward the closer wall could be missed, since the
\v{C}erenkov light could pass through the phototubes without being recorded. 
In our understanding, this is the main reason why the detection
efficiency of Kamiokande-II  never reaches 
100\%, even for the higher energy events, but is at most 
92\%. In order to test this hypothesis, we considered the external part 
of the detector that has a distance from the phototubes smaller 
than their typical distance of 1 meter. 
Since this part has a volume of about 800 m$^3$,
we reproduce the 92\% efficiency if 
$\sim 20$\% of the  events are lost in this way, that is 
a reasonable figure being comfortably smaller than 40\%. 
Again, this effect leads us to expect that the 
loss of signal events is more significant for the 
events produced close to the border, especially
for $f\ge 0.6$, that corresponds to the external part of the 
detector considered just above. We attempt to quantify the 
size of this geometrical 
effect by the function $G[f]=1- (f-0.4)^2 \theta[f-0.4]$,
where we used a quadratic function to describe a smooth behavior 
($\theta[x]$ is the Heaviside function). 
This means that $\sim 8$\% of the total 
number of events are systematically lost  
close to the border, and that even at $f=1$, the loss is 
smaller than 40\%.\\[1ex]
{\em Light attenuation:} The light from the low energy events can 
be lost because of light attenuation, where 
$\lambda\sim 50$ m  (in \cita{hirata}, we read 
that $\lambda$ ``exceeds 50 m at any time''). 
Let us evaluate the average attenuation coefficient for 
Kamiokande-II detector, as a function of the volumetric coordinate
$f$. Considering the simplified case of 
propagation in straight rays, we find the 
differential distribution in the variable $f$ 
by averaging over the other coordinates that specify the position of the event
and the direction of emission:
\begin{equation}
\frac{dA}{df}=\int^1_0\!\! dc\int_0^{2 \pi}\!\! d\phi
\int_0^1\!\! du 
\frac{ u A[u f^{1/3},f^{1/3},c,\phi]+A[f^{1/3},u f^{1/3},c,\phi] }{3\pi}
\label{daf}
\end{equation} 
where $A[r,h,c,\phi]=(e^{-D_{in}/\lambda}+e^{-D_{out}/\lambda})/2$
is the average attenuation for a given direction of observation,
$D_{in}$ and $D_{out}$ are the two distances from the walls of the detector
(see also table~\ref{tab1}),
and we measure the radial coordinate $r$ in units of $R$ and 
the  height $h$ in units of $H/2$. A straightforward calculation
gives the approximation for $dA/df\approx 0.858+0.068\ f$, 
that means that the events produced in the center  
suffer an attenuation effect that is $\sim 10\%$ 
larger. The systematic effect due to light attenuation 
depends on the energy. If we {\em assume} that the signal is 
uniformly distributed in energy,  
we are lead to expect that the loss of low energy 
events from the center is more significant.\\[1ex]
{\em Spectrum:} 
However, the previous point does not take 
into account the real spectrum of the signal. In order to
demonstrate that this is also relevant, 
it is enough to consider a simple one dimensional situation, when
the light propagates just in one direction or the opposite one. 
Let us compare what happens to the events produced in the 
middle ($f\sim 0$) and in the border of the detector ($f\sim 1$). 
In the first case, the light attention effect  
is $e^{-L/2\lambda}\approx 0.87$, where $L= 14$ m 
is the linear size of the detector; 
in the second case, this depends on whether the event propagates toward 
the closer border (no attenuation) or toward the opposite border,
when we have $e^{-L/\lambda}\approx 0.76$.
In the first case, we lose all the events between 
$N_{min}$ and $1.15 N_{min} $, where $N_{min}$ is the minimal
number of photons that gives a trigger; in the second one, we lose half 
of the events between $N_{min}$ and $1.3 N_{min}$ 
(where $e^{L/\lambda}\approx 1.3$). 
If the signal is uniformly distributed in energy, 
we see that at ${\cal O}(L/\lambda)$ the effect is identical 
in the two cases; thus, 
the effect described by eq.~\ref{daf} 
is purely three dimensional.
But if the distribution of the signal increases 
with the energy, as we expect, the loss of signal events 
is more significant for the events produced 
close to the border. In fact, the events that
fall below the detection threshold  in the first case 
(events produced close to $f\sim 0$) are those that 
belong to the least populated part of the spectrum. 
{\em E.g.}, if we assume that the energy spectrum 
of the signal rises linearly from zero, the total number of remaining events 
in the region $N_{min}-1.3 N_{min}$ is 50\% larger in the first case.
This last effect is presumably the less relevant one.

In summary, various effects can change the distribution of the  
detected signal events: the second effect increases a bit the number of 
expected supernova neutrino events close to the border, the other 
two diminish it. It is not possible to estimate precisely the 
systematics deviations from uniformity without a detailed simulation 
of the detector response that takes into account the 
energy distribution of the signal, the role of 
energy fluctuations, {\em etc}.
However, if we account for the two systematics effects that work 
in opposite directions simply by multipling $G[f]$ by $dA/df[f]$, we see that 
the geometrical effect is the most important one. This
is the reason why we claim that the GOF with a uniform volume 
distribution underestimates 
a possible problem with events close to the border: in fact, 
when we apply the SCVM test to the 12 events seen by Kamiokande-II 
we get a GOF of 3.4\% rather than the value that we obtain with 
uniform distribution, 4.6\% (see table~\ref{tab2}).
Similarly, when we discard event 6, the value of the GOF 
becomes 1.8\% rather than 2.4\%  (again in table~\ref{tab2}).
As we see, 
the changes due to a more accurate modeling of the detector response 
are not expected to be very large, and are in any case 
smaller than the changes due to the use of alternative 
statistical tests (see next Appendix).

\section{\sf\color{verdon}  The Anderson-Darling test\label{app:cc}}
The Anderson-Darling (AD) test \cita{scvm_1} is used to test if
a sample of data comes from a specific distribution. It is a
modification of the Kolmogorov-Smirnov test 
and gives more weight to the boundary region
than the Smirnov-Cram\`e{}r-Von Mises test.

Consider a set of data $x_1,x_2,... x_n$, arranged in increasing order.
Given the expected  cumulative distribution function $F[x]$,  
we calculate the values in the points $x_j$:
\begin{equation}
F_j=F[x_j]
\end{equation}
Next, consider  
the empirical cumulative distribution function $F_n[x]$,
that counts the number of events below $x$;
thus, $F_n[x]=j/n$ for $F_j\le u < F_{j+1}$,
where we set $F_0=0$ and $F_{n+1}=1$.
The quantitative indicator of the likelihood 
of the given data proposed by Anderson and Darling is:
\begin{equation}
W^2\equiv 
n \left. \int_0^1 (F_n[x]-u)^2\ \psi[u]\ du \ \right |_{F[x]=u}
\end{equation}
where $\psi$ is a positive function on the interval $(0,1)$.
The case when $\psi=1$ gives the 
well-known Smirnov-Cram\`e{}r-Von Mises test,
that somehow resembles the $\chi^2$ statistics:
\begin{equation}
W^2_{\mbox{\small scvm}}= \frac{1}{12n} + 
\sum_{j=1}^{n} \left( F_j -
\frac{j -1/2}{n} \right)^2
\label{scvm}
\end{equation}
Let us pass to the alternative possibilities. 
The traditional (two sided) version of the 
Anderson-Darling test statistic is 
obtained setting $\psi[u]=1/(u (1-u))$, that gives:
\begin{equation}
W^2_{\mbox{\small traditional}}= -n-\sum_{j=1}^{n} \ \frac{2 j -1}{n} \ 
\log[ F_j\ ( 1-F_{n+1-j}  ) ] 
\label{adtrad}
\end{equation}
This provides a particular sensitivity close to the 
points $u=0$ and $u=1$. A trivial modification consists 
in setting $\psi[u]=1/(1-u)$: in this way, the test becomes particularly 
sensitive close to $u=1$. The modified (or one sided) 
Anderson-Darling test statistic can be written as:
\begin{equation}
W^2_{\mbox{\small modified}}
=
\frac{n}{2} - 2 \sum_{j=1}^{n} F_j-
\sum_{j=1}^{n}  \frac{2 j -1}{n} \ 
 \log[ 1-F_{n+1-j} ] 
\label{admod}
\end{equation}
Since $1/(u(1-u))=1/u+1/(1-u)$, equation~\ref{adtrad} can 
be recovered by summing a third test statistic 
when we replace $\psi[u]=1/u$, that would be suited to  test for 
features close to $u=0$ (the practical formula for this 
third test statistic is obtained replacing $F_j\to 1-F_{n+1-j}$ 
in equation~\ref{admod}).

A null-hypothesis that we want to test is whether a 
pre-selected subset of the KII data is uniformly 
distributed in the detector, as expected if they 
are due to supernova neutrino events.
Thus, in our case the number of data is $n\le 12$,
the coordinate
$x$ is what is called $f$ in Sect.~\ref{sd}, 
and the 
cumulative distribution function is simply $F[x]=x$.
The AD tests of Eqs.~\ref{adtrad} and \ref{admod}
are of particular interest since the points close to 
$f= 1$ are in the border of the volume used in 
the analysis, and, there, it is not possible
to exclude {\em a priori} an effect of contamination
from the background.

Given the test statistics, we 
calculate the value $W^{2}_*$ for the 
subset of KII data in which we are interested.
How often this value results from a uniform distribution?
This can be obtained with a straightforward procedure:
1)~we extract randomly $n$ data from a uniform
distribution for $N$ times ($N=10^8$ in our case);
2)~we calculate the value of $W^2$ for each data set;
3)~finally, we test the frequency of the condition 
$W^2>W^2_*$, namely, the number of times $N_*$
that this condition is satisfied over 
the number of extractions~$N$.

{\em E.g.}, we infer from the test of equation~\ref{adtrad} that 
when all 12 events (resp., all except event 6)
are supposed to be due to supernova neutrinos, we get $W^2_*=4.95$
(resp., $W^2_*=5.80$). A value this high can occur 
as a result of a statistical fluctuations, but 
only in 0.32 \% (resp., 0.13 \%) of the cases; 
this means that we can reject the hypothesis 
that the 12 (resp., 11) events are  
uniformly distributed at the 2.9 
$\sigma$  (resp., 3.2~$\sigma$) significance level.
Similarly, 
we infer from the test of equation~\ref{admod} that 
the case when all events (resp., all except event 6)
are assumed to be signal 
can be rejected at the $3.5$ $\sigma$ 
(resp., 3.8 $\sigma$) significance level.

In principle, it is possible to object 
that the situations when different statistical 
tests give very different outcomes should be regarded as dubious and
that the statistical criterion would be better selected {\em a priori} 
rather than {\em a posteriori}; 
these are the reasons why in this paper 
we do not use exclusively the 
(more powerful) AD test, 
but also the (more conventional) SCVM test.

%% \begin{table}[t]
%% \footnotesize
%% \begin{tabular}{c|cccccccccccccccc}
%% $i$&  1&2&3&4&5&6&7&8&9&10&11&12&13&14&15&16 \\  
%% \hline \\[-2ex]
%% $B_i^{LL}$& 1.6$_5$&1.9$_3$&2.9$_2$&1.2$_2$&2.1$_3$&3.7$_2$&4.5$_5$&8.2$_5$&1.5$_5$&1.5$_2$&1.9$_3$&1.6$_2$&3.8$_2$&2.9$_2$&2.8$_2$&3.8$_2$ \\
%% \hline \\[-2ex]
%% $B_i$& 1.0$_5$&5.4$_4$&3.1$_2$&8.5$_3$&5.3$_4$&7.1$_2$&5.0$_6$&1.0$_5$&1.0$_5$&1.8$_2$&4.0$_4$&1.4$_2$&7.3$_2$&5.2$_2$&1.8$_2$&7.3$_2$ \\
%% \hline \\[-2ex]
%% $E_i$& 20.0&13.5&7.5&9.2&12.8&6.3&35.4&21.0&19.8&8.6&13.0&8.9&6.5&5.4&4.6&6.5 
%% \end{tabular}
%% \caption{\em $1^{st}$ line, the progressive event number; next,
%% the estimated background rate in $\rm {Hz/MeV}$
%% according to the procedure of the `convolution integrals' followed by 
%% LL ($2^{nd}$ line) and the procedure we
%% describe in the text ($3^{rd}$ line); 
%% $4^{th}$ line, the energy of the event in {\rm MeV}. A subscript 
%% indicates the exponent: e.g., $3_2$ means $3\cdot 10^{-2}$.}
%% \endt{able}

%\newpage

~~~~~~~~~~~~~~~~~~~~~~~~~~~~~~~~~~~~~~~~~~~~~~~~~~~~~~~~~~~~~~~~~~~
\begin{twocolumn}

\section*{\sf\color{verdon}  References}

\footnotesize 

The references are listed in order of time within 
6 topics: SN1987A data \cita{hirataPRL,imb,baksan,mb}; 
other relevant information on Kamiokande-II \cita{hirata,moriond,hirataPHD};
theoretical works on supernova neutrinos 
\cita{nad,del1,del2,latti,nadd,bahcall,keil,ale,ir,marco,del3};
(an incomplete list of) 
analyses of SN1987A data \cita{raffs,am,lambLoredo,lambloredoPRIV,
aldo,dr,mira,luna}; statistics \cita{scvm_1,scvm_2,ad};
talks at three recent conferences on SN1987A \cita{webs}.

\def\refname{

\end{twocolumn}


\vskip-1cm}
\baselineskip=1.15em


\begin{thebibliography}{99}

%\setcounter{enumiv}{-1}
%\bibitem{}

%%% Data

\bibitem{hirataPRL}
K~Hirata {\it et al } [Kamiokande-II Collaboration],
%``Observation Of A Neutrino Burst From The Supernova Sn1987a,''
Phys.\ Rev.\ Lett.\ {\bf 58} (1987) 1490.
%%CITATION = PRLTA,58,1490;%%

\bibitem{imb}
R M~Bionta {\it et al.},
% ``OBSERVATION OF A NEUTRINO BURST IN COINCIDENCE WITH SUPERNOVA SN1987A IN
%THE LARGE MAGELLANIC CLOUD,''
Phys.\ Rev.\ Lett.\ {\bf 58} (1987) 1494.
%%CITATION = PRLTA,58,1494;%%

\bibitem{baksan}
E N~Alekseev, L N~Alekseeva, I V~Krivosheina and V I~Volchenko,
%``DETECTION OF THE NEUTRINO SIGNAL FROM SN1987A IN THE LMC USING THE INR
%BAKSAN UNDERGROUND SCINTILLATION TELESCOPE,''
Phys.\ Lett.\ B {\bf 205}, 209 (1988).
%%CITATION = PHLTA,B205,209;%%

\bibitem{mb}
M~Aglietta {\it et al.},
   %``On the event observed in the Mont Blanc Underground Neutrino observatory
  %during the occurrence of Supernova 1987a,''
  Europhys.\ Lett.\  {\bf 3} (1987) 1315.
  %%CITATION = EULEE,3,1315;%%



%%% More on Kamiokande II

\bibitem{hirata}
K S~Hirata {\it et al.},
%``OBSERVATION IN THE KAMIOKANDE-II DETECTOR OF THE NEUTRINO BURST FROM
%SUPERNOVA SN1987A,''
Phys.\ Rev.\ D {\bf 38} (1988) 448.
 %%CITATION = PHRVA,D38,448;%%

\bibitem{moriond} 	
K S~Hirata {\em et al.}, in
``The Standard model: supernova 1987A'' 
proceedings of leptonic session of  
XXII$^{nd}$ Rencontre de Moriond, Les Arcs, March 1988
edited by J.~Tran Thanh Van, p. 689.
%%CITATION = C87/07/27;%%


\bibitem{hirataPHD}
K S~Hirata,
``Search for supernova neutrinos at Kamiokande-II'',
Tokyo Univ. Ph.D. Thesis, ICRR-239-91-08.



%%% Theory

\bibitem{nad}
D K~Nadyozhin,
Astrophys.\ Space Sci.\  {\bf 53} (1978) 131. 

\bibitem{del1}
J R~Wilson  in {\em Numerical Astrophysics}, 
eds.\ J~Centrella, J~LeBlanc, R L~Bowers, 
page 422 (Jones and Bartlett, Boston, 1985).

\bibitem{del2}
H A~Bethe and J R~Wilson,
Astrophys.\ J.\  {\bf 295} (1985) 14. 

\bibitem{latti}
J M  Lattimer, A Yahil, 
Astrophys.\ J.\ {\bf 340} (1989) 426.

\bibitem{bahcall} J N~Bahcall, cap~15 of 
{\em Neutrino astrophysics,} Cambridge University Press, 1989.

\bibitem{nadd}
D K Nadyozhin
%''The neutrino signal from a collapsing star''
in {\em Supernovae},  
eds.~S A~Bludman, R Mochkovitch, J Zinn-Justin, 
page 301  (Les Houches session LIV, 1990, 
North Holland \& Elsevier, 1994). 



\bibitem{keil}
M T~Keil, G G~Raffelt and H T~Janka,
%``Monte Carlo study of supernova neutrino spectra formation,''
Astrophys.\ J.\ {\bf 590} (2003) 971
[arXiv:astro-ph/0208035].
%%CITATION = ASTRO-PH 0208035;%%


\bibitem{ale} A~Strumia and F~Vissani,
%``Precise quasielastic neutrino nucleon cross section,''
Phys.\ Lett.\ B {\bf 564} (2003) 42
[arXiv:astro-ph/0302055].
%%CITATION = ASTRO-PH 0302055;%%


\bibitem{ir}
V S Imshennik and O G Ryazhskaya,
%``A rotating collapsar and possible interpretation of the LSD neutrino
%signal from SN 1987A,''
Astron.\ Lett.\  {\bf 30} (2004) 14.
%%CITATION = ALETE,30,14;%%

\bibitem{marco}

N~Y~Agafonova {\it et al.},
%``Study of the effect of neutrino oscillations on the supernova neutrino
%signal in the LVD detector,''
Astropart.\ Phys.\  {\bf 27} (2007) 254
[arXiv:hep-ph/0609305].
%%CITATION = APHYE,27,254;%%



\bibitem{del3}
H Th~Janka, K~Langanke,
A~Marek, G~Mart\'\i{}nez-Pinedo, B~M\"{u}ller,
submitted to the {\em Bethe Centennial Volume} of 
Physics Reports [astro-ph/0612072].





%%% Analyses

\bibitem{raffs}
B~Jegerlehner, F~Neubig and G~Raffelt,
  %``Neutrino Oscillations and the Supernova 1987A Signal,''
Phys.\ Rev.\ D {\bf 54} (1996) 1194
[arXiv:astro-ph/9601111].
%%CITATION = ASTRO-PH 9601111;%%

\bibitem{am}
A S Mal'gin,
%Analysis of integral and averaged characteristics of the IMB and Kamioka signals from SN1987A
Nuovo Cim.\ C {\bf 21} (1998) 317. 
%%CITATION = NUCIA,C21,317;%%

\bibitem{lambLoredo}
T J~Loredo and D Q~Lamb,
%``Bayesian analysis of neutrinos observed from supernova SN 1987A,''
Phys.\ Rev.\ D {\bf 65} (2002) 063002
[arXiv:astro-ph/0107260].
%%CITATION = ASTRO-PH 0107260;%%

\bibitem{lambloredoPRIV}
Y~Totsuka, A~Mann, S -B~Kim as quoted in previous paper.

\bibitem{aldo}
M L~Costantini, A~Ianni and F~Vissani,
%``SN1987A and the properties of neutrino burst,''
Phys.\ Rev.\ D {\bf 70} (2004) 043006
[arXiv:astro-ph/0403436].
%%CITATION = ASTRO-PH 0403436;%%


\bibitem{dr}
D K Nadyozhin and V S Imshennik, %``Physics of Supernovae,''
Int.\ J.\ Mod.\ Phys.\  A {\bf 20} (2005) 6597. 
%%CITATION = IMPAE,A20,6597;%%

\bibitem{mira}
A~Mirizzi and G G~Raffelt,
%``New analysis of the SN 1987A neutrinos with a flexible spectral shape,''
Phys.\ Rev.\ D {\bf 72} (2005) 063001
[arXiv:astro-ph/0508612].
%%CITATION = ASTRO-PH 0508612;%%

\bibitem{luna}
C~Lunardini,
%``The diffuse supernova neutrino flux, star formation rate and SN1987A,''
Astropart.\ Phys.\  {\bf 26} (2006) 190
[arXiv:astro-ph/0509233].
%%CITATION = ASTRO-PH 0509233;%%





%%% STATISTICS


\bibitem{scvm_1}
M A Stephens,
Journal of the American Statistical Association 69 (1974) 730.

\bibitem{scvm_2}
S  Csorgo, J J  Faraway,
Journal of the Royal Statistical Society B 58 (1996) 221.

\bibitem{ad}
T W  Anderson, D A  Darling,
American Statistical Association Journal 49 (1954) 765.


\bibitem{webs}
Web sites the conferences held in Moscow:
http:/$\!$/lvd.ras.ru/SN1987A/en/materials/ind\-ex.php;
Hawaii, http:/$\!$/sn1987a-20th.physics.\-uci.edu/Program1.htm; 
Venice, http:/$\!$/neutri\-no.pd.infn.it/conference2007/talks.html.





\end{thebibliography}
\end{document}